\documentclass[aps,graphicx,12pt,showkeys,showpacs,onecolumn]{revtex4}
\usepackage{amsmath}
\usepackage{amscd}
\usepackage{graphicx}
\usepackage{subfigure}
\usepackage{appendix}

\begin{document}

\title[Short Title]{Efficient shortcuts to adiabatic passage for the fast populations transfer in multiparticle systems}

\author{Ye-Hong Chen$^{1}$}
\author{Qing-Qin Chen$^{2}$}
\author{Yan Xia$^{1,}$\footnote{E-mail: xia-208@163.com}}
\author{Jie Song$^{3,}$\footnote{E-mail: jsong@hit.edu.cn}}

\affiliation{$^{1}$Department of Physics, Fuzhou University, Fuzhou
350002, China\\$^{2}$Zhicheng College, Fuzhou University, Fuzhou
350002, China\\$^{3}$Department of Physics, Harbin Institute of
Technology, Harbin 150001, China}


\begin{abstract}
Achieving fast population transfer (FPT) in multiparticle systems
based on the cavity quantum electronic dynamics is an
outstanding challenge. In this paper, motivated by the quantum Zeno
dynamics, a shortcut for performing the FPT of ground states
in multiparticle systems with the invariant based inverse
engineering is proposed.  Numerical simulation demonstrates that a
perfect population transfer of ground states in multiparticle
systems can be rapidly achieved in one step, and the FPT is robust
to both the cavity decay and atomic spontaneous emission.
Additionally, this scheme is not only implemented without requiring
extra complex conditions, but also insensitive to variations of the
parameters.

\end{abstract}


\maketitle
\section{INTRODUCTION}
Reliable population transfer of a quantum system with
time-dependent interacting fields has become a significant
ingredient in the quantum information processing for various
applications ranging from quantum storage to quantum communication
\cite{JLMPMSKSBPrl06,KBHTBWSRmp98,PKITMSRmp07,NVVTHBWSKBArpc01}. It
has already drawn great attention in recent years
\cite{CPYPra10,MATSGHRJPra05}. Several approaches have been proposed
for attaining complete population transfers with different methods,
including $\pi$ pulses, composite pulses, rapid adiabatic passage
(RAP), stimulated Raman adiabatic passage (STIRAP), and their
variants \cite{KBHTBWSRmp98,PKITMSRmp07,NVVTHBWSKBArpc01}. However,
most of them have some shortcomings, say, $\pi$ pulses is fast yet
highly sensitive to variations in the pulse area, and to
inhomogeneities in the sample \cite{ARXDAJGMNjp12}, the adiabatic passage technique is
robust versus variations in the experimental parameters while it usually
needs a relatively long interaction time. If the required evolution
time is too long, the scheme may be useless, because decoherence
would spoil the intended dynamics. Therefore, accelerating the
dynamics towards the perfect final outcome is a good idea and
perhaps the most reasonable way to actually fight against the
decoherence that is accumulated during a long operation time.

Recently, a lot of work has been done in finding shortcuts to
adiabaticity for the two- or three-level atomic system
\cite{MDSARJpca03,MDSARJcp08,XCILARDGJGMPrl10,SMNKPrsla10,MVBJpamt09,XCJGMPra12,XCETJGMPra11,MLYXLTSJSNBAPra14}.
By means of resonant laser pulses, Chen and Muga have successfully
performed fast population transfer (FPT) in three-level systems via invariant-based inverse
engineering \cite{XCJGMPra12}. A quantum computation network has
long been thought to partition into a sequence of
one-qubit rotations and two-qubit gates \cite{TSHWPrl95}.
Nevertheless, it is too large to construct a quantum computation network to
perform computation by decomposing into one-qubit rotations and
two-qubit gates. So, the FPT in a multiparticle system is a
fundamental operation for scalable quantum information processors.
However, it is a pity that most of the previous studies based on the
invariant-based inverse engineering for achieving FPTs are in two-
or three-level single-atom systems, and it is very hard to directly
design a model for the FPT in multiparticle systems. Until recently,
Lu \emph{et al.} have proposed a scheme to implement the quantum
state transfer between two three-level atoms based on the
invariant-based inverse engineering in the cavity quantum electronic
dynamics (QED) system \cite{MLLTSYXJS13}. They sent two atoms
through the cavity with a short time interval, and the atoms
suffered the oppositive variation tendency in the time-dependent
laser pulse and atom-cavity coupling. Through designing
related parameters and controlling the time interval between the two
atoms sent through the cavity, they effectively implemented
ultrafast quantum state transfer between two $\Lambda$-type atoms.
Reference \cite{MLLTSYXJS13} successfully
introduced shortcuts to adiabatic passage into cavity QED systems.
However, Ref. \cite{MLLTSYXJS13} is limited
by the following:
(i) Only quantum state transfer in a two-atom system could be
realized. When it comes to more complex systems, for example,
multiparticle systems, cavity coupling systems, and
cavity-fiber-atom combined systems, this scheme is useless; new
designs are required in a different situation. (ii) Sequential
operations were needed in a two atoms system; this may eliminate the
possibility of success in experiment.

On the other hand, the quantum Zeno effect which has been tested in many
experiments is the inhibition of transitions between quantum states by
frequent measurements \cite{BMECGSJmp77,WMIDJHJJBDJWPra90,PWHWTHAZMAKPrl95,RJCPst}.
The system can actually evolve away from its initial state while it still
remains in the so-called Zeno subspace determined by the measurement
when frequently projected onto a multi-dimensional subspace.
This was called ``quantum Zeno dynamics'' by Facchi and Pascazio in 2002 \cite{PFSPPrl02}.
And quantum Zeno dynamics (QZD) can be achieved via continuous coupling between
the system and an external system instead of discontinuous measurements.
In general, we assume that a dynamical evolution process is governed by
the Hamiltonian $H_{K}=H_{obs}+KH_{meas}$, where $H_{obs}$ is Hamiltonian of the quantum system
investigated, $K$ is a coupling constant, and $H_{meas}$ is viewed as an
additional interaction Hamiltonian performing the measurement. In the limit $K\rightarrow\infty$
the system will remain in the same Zeno subspace as that of its initial state.
The evolution operator is described as
$U(t)=\exp(-it\sum_{n}K\eta_{n}P_{n}+P_{n}H_{obs}P_{n})$, with $P_{n}$ being the eigenvalue projection
of $H_{meas}$ with eigenvalues $\eta_{n}$ ($H_{meas}=\sum_{n}\eta_{n}P_{n}$).

To more widely generalize the efficiency and application of the FPT
in multiparticle systems based on shortcuts to adiabatic passage in
cavity QED systems, motivated by the space division of QZD, we
propose an effective method by invariant-based inverse engineering.
Compared with previous works, this protocol has the following
advantages. First, the fast population transfer in a
multiparticle system can be achieved in
one step. Secondly, the shortcut to the adiabatic passage is reliable for
dealing with much more complex situations, for example, multiparticle
systems, cavity coupling systems, and cavity-fiber-atom combined
systems.

The paper is structured as follows. In Sec. \textrm{II}, we
construct a shortcut passage for FPT in a system with two
$\Lambda$-type atoms trapped in a cavity. A resonant time-dependent
laser pulse and a resonant ordinary atom-cavity coupling are applied
to each atom. In Sec. \textrm{III}, we analyze the feasibility of
the FPT in multiparticle systems based on the shortcut proposed in
Sec. \textrm{II}. Sec. \textrm{IV} is the conclusion.

\section{Shortcuts to adiabatic passage for the fast populations transfer in two-atom system}
\begin{figure}
 \scalebox{0.45}{\includegraphics {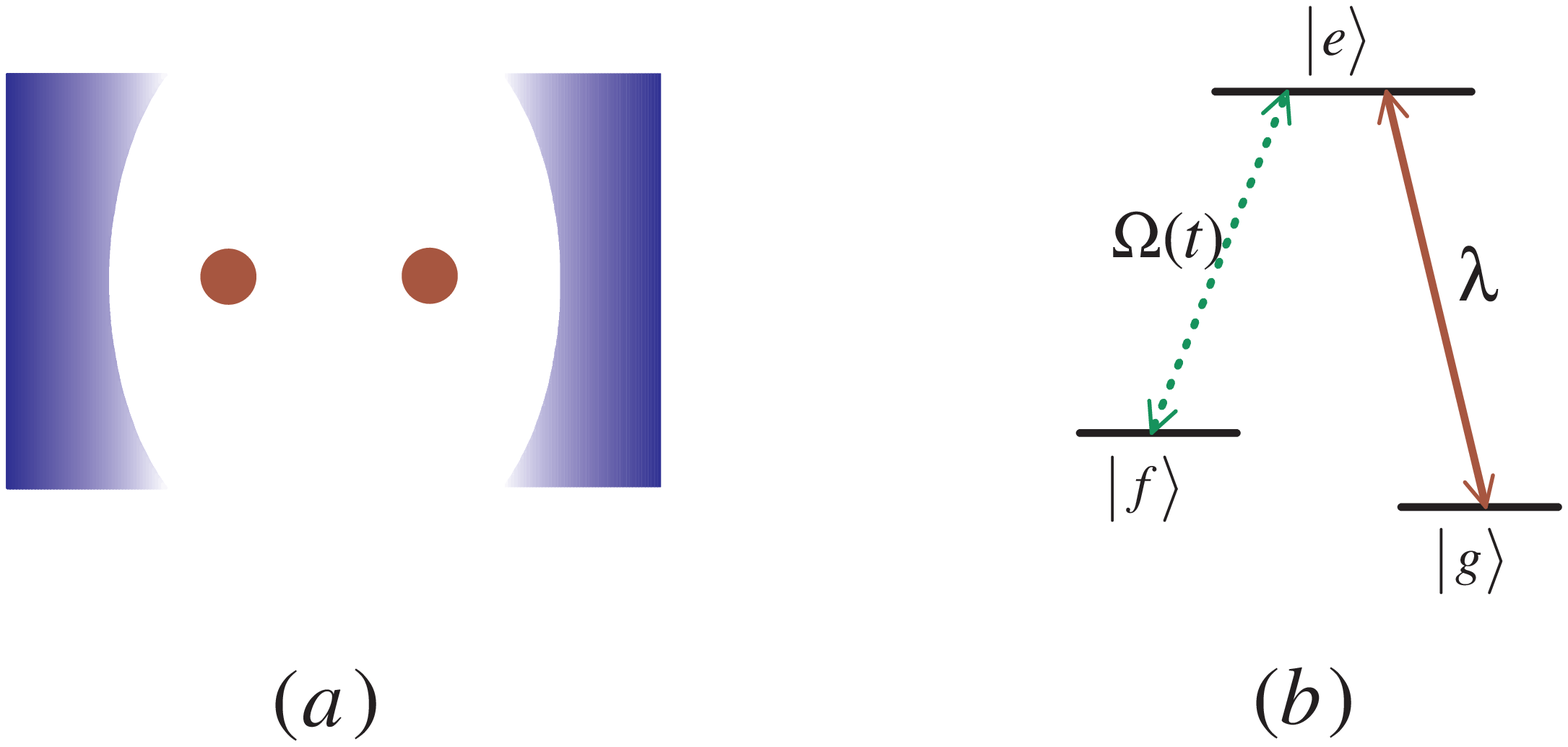}}
 \caption{(a) Cavity-atom combined system. (b) Atomic level configuration.}
 \label{model}
\end{figure}

As shown in Fis. \ref{model}, we consider that two $\Lambda$-type
atoms 1 and 2 are trapped in a cavity $c$. Each atom has an excited
state $|e\rangle$ and two ground states $|f\rangle$ and $|g\rangle$.
The atomic transition $|f\rangle\leftrightarrow|e\rangle$  is
resonantly driven through a time-dependent laser pulse with Rabi
frequency $\Omega(t)$, and the transition
$|g\rangle\leftrightarrow|e\rangle$ is resonantly coupled to the
cavity mode with coupling constant $\lambda$. The whole Hamiltonian
in the interaction picture is written as
\begin{eqnarray}\label{eq1}
  H_{I}   &=&H_{al}+H_{ac},                                         \cr\cr
  H_{al}&=&\sum_{k=1,2}{\Omega_{k}(t)|e\rangle_{k}\langle f|+H.c.}, \cr\cr
  H_{ac}&=&\sum_{k=1,2}{\lambda_{k}|e\rangle_{k}\langle g|a+H.c.},
\end{eqnarray}
where subscript $k$ denotes the $k$th atom, and $a$ is the annihilation operator for the cavity.
If the initial state is $|\psi_{0}\rangle=-|f\rangle_{1}|g\rangle_{2}|0\rangle_{c}$, the whole system evolves
in the subspace spanned by
\begin{eqnarray}\label{eq2}
  |\psi_{1}\rangle&=&|f\rangle_{1}|g\rangle_{2}|0\rangle_{c},   \cr
  |\psi_{2}\rangle&=&|e\rangle_{1}|g\rangle_{2}|0\rangle_{c},   \cr
  |\psi_{3}\rangle&=&|g\rangle_{1}|g\rangle_{2}|1\rangle_{c},   \cr
  |\psi_{4}\rangle&=&|g\rangle_{1}|e\rangle_{2}|0\rangle_{c},   \cr
  |\psi_{5}\rangle&=&|g\rangle_{1}|f\rangle_{2}|0\rangle_{c}.
\end{eqnarray}
In light of QZD, we rewrite the Hamiltonian in Eq. (\ref{eq1})
with the eigenvectors of $H_{ac}$ (we set $\lambda_{1}=\lambda_{2}=\lambda$),
\begin{eqnarray}\label{eq3}
  |\phi_{1}\rangle&=&\frac{1}{\sqrt{2}}(-|\psi_{2}\rangle+|\psi_{4}\rangle),                   \cr
  |\phi_{2}\rangle&=&\frac{1}{2}(|\psi_{2}\rangle+\sqrt{2}|\psi_{3}\rangle+|\psi_{4}\rangle),  \cr
  |\phi_{3}\rangle&=&\frac{1}{2}(|\psi_{2}\rangle-\sqrt{2}|\psi_{3}\rangle+|\psi_{4}\rangle),
\end{eqnarray}
with eigenvalues $E_1=0$, $E_2=\sqrt{2}\lambda$, and $E_3=-\sqrt{2}\lambda$. We obtain
\begin{eqnarray}\label{eq4}
  H'_{I}&=&H'_{al}+H'_{ac},                                                           \cr\cr
  H'_{ac}&=&\sum_{n=1}^{3}{E_{n}|\phi_{n}\rangle\langle\phi_{n}|},                    \cr\cr
  H'_{al}&=&\frac{\Omega_{1}(t)}{\sqrt{2}}(-|\phi_{1}\rangle\langle\psi_{1}|)
           +\frac{\Omega_{1}(t)}{2}|\phi_{2}\rangle\langle\psi_{1}|
           +\frac{\Omega_{1}(t)}{2}|\phi_{3}\rangle\langle\psi_{1}|
           +\frac{\Omega_{2}(t)}{\sqrt{2}}|\phi_{1}\rangle\langle\psi_{5}|            \cr\cr
         &&+\frac{\Omega_{2}(t)}{2}|\phi_{2}\rangle\langle\psi_{5}|
           +\frac{\Omega_{2}(t)}{2}|\phi_{3}\rangle\langle\psi_{5}|
           +H.c..
\end{eqnarray}
It is obvious that there are four non-zero energy eigenvalues $\pm\Omega_{1}(t)$ and $\pm\Omega_{2}(t)$ for the Hamiltonian $H'_{al}$.
Therefore, setting $\sqrt{2}\lambda\gg \Omega_{k}(t)$, the condition $H'_{ac}\gg H'_{al}$ and
the Zeno condition $K\rightarrow\infty$ are satisfied ($H'_{al}$ and $H'_{ac}$ correspond to $H_{obs}$ and $KH_{meas}$ in
Sec. I, respectively). Performing the unitary transformation $U=e^{-iH'_{ac}t}$ under condition $H'_{ac}\gg H'_{al}$, we obtain
\begin{eqnarray}\label{eq5}
  H^{eff}_{al}&=&\frac{\Omega_{1}(t)}{\sqrt{2}}(-|\phi_{1}\rangle\langle\psi_{1}|)
                 +\frac{\Omega_{1}(t)}{2}(e^{i\sqrt{2}\lambda t}|\phi_{2}\rangle\langle\psi_{1}|)
                 +\frac{\Omega_{1}(t)}{2}(e^{-i\sqrt{2}\lambda t}|\phi_{3}\rangle\langle\psi_{1}|)
                 +\frac{\Omega_{2}(t)}{\sqrt{2}}(|\phi_{1}\rangle\langle\psi_{5}|)                       \cr\cr
               &&+\frac{\Omega_{2}(t)}{2}(e^{i\sqrt{2}\lambda t}|\phi_{2}\rangle\langle\psi_{5}|)
                 +\frac{\Omega_{2}(t)}{2}(e^{-i\sqrt{2}\lambda t}|\phi_{3}\rangle\langle\psi_{5}|)
                 +H.c..
\end{eqnarray}
The terms with the oscillating frequency $\sqrt{2}\lambda$ are possible to be ignored in the present case.
And the Hilbert subspace is split into three invariant Zeno subspaces $H_{p0}=\{|\psi_{1}\rangle,|\psi_{5}\rangle,|\phi_{1}\rangle\}$,
$H_{p1}=\{|\phi_{2}\rangle\}$, and $H_{p2}=\{|\phi_{3}\rangle\}$.

The above analysis provides a classical space division via QZD. Nevertheless, it is easily found from Eq. \ref{eq5} that the transition
$|\psi_{1}\rangle\leftrightarrow|\phi_{2}\rangle(|\phi_{3}\rangle)\leftrightarrow|\psi_{5}\rangle$ is still difficult to realize even when $\Omega_{k}(t)$ is very close to
$\lambda$. Therefore, we assume $\sqrt{2}\lambda$ is slightly
larger than $\Omega_{k}$ and divide the system into three subsystems,
\begin{eqnarray}\label{eq6}
  S_{1}=\{|\psi_{1}\rangle,\ |\phi_{2}\rangle,\ |\psi_{5}\rangle \}, \
  S_{2}=\{|\psi_{1}\rangle,\ |\phi_{1}\rangle,\ |\psi_{5}\rangle \}, \
  S_{3}=\{|\psi_{1}\rangle,\ |\phi_{3}\rangle,\ |\psi_{5}\rangle \}.
\end{eqnarray}
We neglect the interaction between the states in each of the
subsystems $S_{1}$ and $S_{3}$ for the moment since the interaction
is far weaker than that in subsystem $S_{2}$. Then the system can be
considered as a three-level single-atom system with two ground
states $|\psi_{1}\rangle$ and $|\psi_{5}\rangle$ and an excited
state $|\phi_{1}\rangle$. If we replace $|\psi_{1}\rangle$ as
$|\psi_{0}\rangle$, the Hamiltonian for STIRAP reads
\begin{eqnarray}\label{eq7}
  H_{S_2}(t)=\frac{1}{\sqrt{2}}\left(
                                    \begin{array}{ccc}
                                        0 & \Omega_{1}(t) & 0             \\
                                        \Omega_{1}(t) & 0 & \Omega_{2}(t) \\
                                        0 & \Omega_{2}(t) & 0             \\
                                    \end{array}
                               \right).
\end{eqnarray}
The corresponding instantaneous eigenstates $|\Phi_{n}\rangle$,
with eigenvalues $\eta_{0}=0$ and $\eta_{\pm}=\pm\chi/\sqrt{2}$, with
$\chi=\sqrt{\Omega_1^2(t)+\Omega_{2}^2(t)}$ and $\theta=\arctan[\Omega_{1}(t)/\Omega_{2}(t)]$, are
\begin{eqnarray}\label{eq8}
  |\Phi_{0}(t)\rangle=
    \left(
     \begin{array}{c}
       \cos{\theta}  \\
       0             \\
       -\sin{\theta} \\
     \end{array}
    \right),    \
  |\Phi_{\pm}(t)\rangle=
    \frac{1}{\sqrt{2}}
    \left(
     \begin{array}{c}
       \sin{\theta}  \\
       \pm1          \\
       \cos{\theta}  \\
     \end{array}
    \right).
\end{eqnarray}
Population transfer from the initial state $|\psi_{0}\rangle$ to
the state $|\psi_{5}\rangle$ is achieved adiabatically
along the dark state $|\Phi_{0}\rangle$ when the adiabatic
condition $|\dot{\theta}|\ll |\frac{1}{\sqrt{2}}\chi|$ is
satisfied. To speed up the transfer by using the dynamics of
invariant-based inverse engineering, we need to introduce an invariant Hermitian operator $I_{S_{2}}(t)$, which
satisfies $i\partial I_{S_{2}}(t)/\partial t=[H_{S_{2}}(t),I_{S_{2}}(t)]$
\cite{XCILARDGJGMPrl10,XCETJGMPra11,XCJGMPra12,YZLJQLHJWMKJGZPra96,YZLJQLHJWMKJGZJmp96}, for $H_{S_{2}}(t)$ possesses the SU(2) dynamical symmetry. And $I_{S_{2}}(t)$ is given by
\begin{eqnarray}\label{eq9}
   I_{S_{2}}(t)=\frac{1}{\sqrt{2}}\chi
              \left(
                \begin{array}{ccc}
                 0 & \cos{\gamma}\sin{\beta} & -i\sin{\gamma}           \\
                 \cos{\gamma}\sin{\beta} & 0 & \cos{\gamma}\cos{\beta}  \\
                 i\sin{\gamma} & \cos{\gamma}\cos{\beta} & 0            \\
              \end{array}
             \right),
\end{eqnarray}
the time-dependent auxiliary parameters $\gamma$ and $\beta$ satisfy the equations
\begin{eqnarray}\label{eq10}
  \dot{\gamma}&=&\frac{1}{\sqrt{2}}(\Omega_{1}\cos{\beta}-\Omega_{2}\sin{\beta}),           \cr
  \dot{\beta}&=&\frac{1}{\sqrt{2}}\tan{\gamma}(\Omega_{2}\cos{\beta}+\Omega_{1}\sin{\beta}),
\end{eqnarray}
where the dot represents a time derivative. By inversely deriving from eq. (\ref{eq10}), the explicit
expressions of $\Omega_{1}(t)$ and $\Omega_{2}(t)$ are as follows:
\begin{eqnarray}\label{eq11}
  \Omega_{1}(t)=\sqrt{2}(\dot{\beta}\cot{\gamma}\sin{\beta}+\dot{\gamma}\cos{\beta}), \cr
  \Omega_{2}(t)=\sqrt{2}(\dot{\beta}\cot{\gamma}\cos{\beta}-\dot{\gamma}\sin{\beta}).
\end{eqnarray}
The eigenstates $|\Psi_{n}\rangle$ of
the invariant $I_{S_{2}}(t)$, with eigenvalues $\varepsilon_{0}=0$ and $\varepsilon_{\pm}=\pm1$, are
\begin{eqnarray}\label{eq12}
  |\Psi_{0}(t)\rangle=
    \left(
     \begin{array}{c}
       \cos{\gamma}\cos{\beta}      \\
       -i\sin{\gamma}               \\
       -\cos{\gamma}\sin{\beta}     \\
     \end{array}
    \right),   \
  |\Psi_{\pm}(t)\rangle=
    \frac{1}{\sqrt{2}}
    \left(
     \begin{array}{c}
       \sin{\gamma}\cos{\beta}\pm i\sin{\beta}    \\
       i\cos{\gamma}                              \\
       -\sin{\gamma}\sin{\beta}\pm i\cos{\beta}  \\
     \end{array}
    \right).
\end{eqnarray}
The general solution of the Schr\"{o}dinger equation with respect to the
instantaneous eigenstates of $I_{S_{2}}(t)$ are written as
\begin{eqnarray}\label{eq13}
  |\Psi(t)\rangle=\sum_{m=0,\pm}{C_{m}e^{i\alpha_{m}}|\Psi_{m}(t)\rangle},
\end{eqnarray}
where $C_{m}$ is a time-independent amplitude and $\alpha_{m}$ is the
Lewis-Riesenfeld phase according to Lewis Riesenfeld theory \cite{HRLWBRJmp69}, and
the form of $\alpha_m$ is
\begin{eqnarray}\label{eq14}
  \alpha_{m}(t_{f})=\int_{0}^{t_{f}}dt\langle\Psi_{m}(t)|
                         [i\frac{\partial}{\partial t}-H_{S_{2}}(t)]|\Psi_{m}(t)\rangle,
\end{eqnarray}
where $t_f$ is the total interaction time. Similarly, in our case $\alpha_{0}=0$, and
\begin{eqnarray}\label{eq15}
  \alpha_{\pm}=\mp\int_{0}^{t_{f}}dt{[\dot{\beta}\sin{\gamma}+\frac{1}
  {\sqrt{2}}(\Omega_{1}\sin{\beta}+\Omega_{2}\cos{\beta})\cos{\gamma}]}.
\end{eqnarray}
In order to get the target state
$|\psi_{5}\rangle$ along the invariant eigenstate $|\Psi_{0}(t)\rangle$, we
suitably choose the feasible parameters $\gamma(t)$ and $\beta(t)$
\begin{eqnarray}\label{eq16}
  \gamma(t)=\epsilon,\ \beta(t)=\pi t/2t_{f},
\end{eqnarray}
where $\epsilon$ is a small value, which satisfies $(\sin{\epsilon})^{-1}=4N$ ($N=1,2,3,\cdots$)
for a high fidelity of the target state \cite{XCJGMPra12}. And we obtain
\begin{eqnarray}\label{eq17}
  \Omega_{1}(t)=(\pi/\sqrt{2}t_{f})\cot{\epsilon}\sin(\pi t/2t_{f}),\cr
  \Omega_{2}(t)=(\pi/\sqrt{2}t_{f})\cot{\epsilon}\cos(\pi t/2t_{f}).
\end{eqnarray}

Once the Rabi frequencies are specially designed, the FPT of the states
in subsystem $S_{2}$ will be implemented. Afterwards, we analyze the
population transfer of the states in subsystems $S_{1}$ and $S_{3}$. Analyzing the population transfer in these two subsystems, by contrast, the whole system must be taken into consideration rather than only the subsystem. We consequently introduce two vectors
$|\mu_{1}\rangle=\frac{1}{\sqrt{2}}(|\phi_{2}\rangle-|\phi_{3}\rangle)=|\psi_{3}\rangle$ and
$|\mu_{2}\rangle=\frac{1}{\sqrt{2}}(|\phi_{2}\rangle+|\phi_{3}\rangle)=\frac{1}{\sqrt{2}}(|\psi_{2}\rangle+|\psi_{4}\rangle)$
for rewriting the Hamiltonian in Eq. (\ref{eq4}). We have
\begin{eqnarray}\label{eq 117}
  H_{re}&=&\frac{1}{\sqrt{2}}\Omega_{1}(t)(-|\psi_{1}\rangle\langle\phi_{1}|)
          +\frac{1}{\sqrt{2}}\Omega_{2}(t)|\psi_{5}\rangle\langle\phi_{1}|
          +\frac{1}{\sqrt{2}}\Omega_{1}(t)|\psi_{1}\rangle\langle\mu_{2}|   \cr
        &&+\frac{1}{\sqrt{2}}\Omega_{2}(t)|\psi_{5}\rangle\langle\mu_{2}|
          +\sqrt{2}\lambda|\psi_{3}\rangle\langle\mu_{2}|
          +H.c..
\end{eqnarray}
We find that there is a dark state for the Hamiltonian
$H_{re}$, and the dark state is
\begin{eqnarray}\label{eq 118}
  |Dark\rangle&=&\frac{1}{N_{2}}(\Omega_{2}(t)|\psi_{1}\rangle-\frac{\Omega_{1}(t)\Omega_{2}(t)}
                 {\lambda}|\psi_{3}\rangle+\Omega_{1}(t)|\psi_{5}\rangle)                     \cr
              &=&\frac{1}{N_{2}}[\Omega_{2}(t)|\psi_{1}\rangle-
                  \frac{\Omega_{1}(t)\Omega_{2}(t)}{\sqrt{2}\lambda}(|\phi_{2}\rangle-|\phi_{3}\rangle)
                  +\Omega_{1}(t)|\psi_{5}\rangle]
              ,
\end{eqnarray}
with $N_{2}=\sqrt{\Omega_{1}^{2}+\Omega_{2}^{2}+(\Omega_{1}\Omega_{2}/\lambda)^{2}}$.
The result shows that, based on STIRAP, the states $|\phi_{1}\rangle$
and $|\mu_{2}\rangle$ are neglected when the adiabatic
condition for the whole system is satisfied. However, the adiabatic condition for the whole system
can not be satisfied since we have designed two special Rabi frequencies
$\Omega_{1}$ and $\Omega_{2}$, and we learn from Ref. \cite{XCJGMPra12} that the state
$|\phi_{1}\rangle$ is absolutely populated into a relatively large extent
for speeding up the population transfer. Hence, it is very necessary to analyze
whether the state $|\mu_{2}\rangle$ can still be neglected or not with these two special Rabi frequencies. And the effect of the state $|\mu_{2}\rangle$ during
the evolution of the whole system is worth studying. By solving the characteristic equation of
$H_{re}$, we conclude that the smallest difference between an arbitrary eigenvalue and $0$ is
\begin{eqnarray}\label{eq 119}
  |\Delta E|=\vartheta/\sqrt{2}=\sqrt{\frac{\Omega_{1}^{2}+\Omega_{2}^{2}+2\lambda^{2}-\varpi}{2}},
\end{eqnarray}
with $\varpi=\sqrt{(\Omega_{1}^{2}-\Omega_{2}^{2})^2+4\lambda^{4}}$ and
$\vartheta=\sqrt{\Omega_{1}^{2}+\Omega_{2}^{2}+2\lambda^{2}-\varpi}$, the corresponding eigenstates are
\begin{eqnarray}\label{eq 120}
  |\Theta_{+}\rangle&=& \frac{1}{N_{e}}\{
               \frac{\Omega_{1}}{\varsigma}[\frac{\vartheta^{2}}{2}-(\Omega_{2}^{2}+2\lambda^{2})]|\psi_{1}\rangle
               -\frac{\Omega_{2}}{\varsigma}[\frac{\vartheta^{2}}{2}-(\Omega_{1}^{2}+2\lambda^{2})]|\psi_{5}\rangle          \cr\cr
             &&-\frac{\vartheta({2\lambda^{2}+\varpi})}{2\varsigma}|\phi_{1}\rangle
               -\frac{\vartheta}{2\lambda}|\mu_{2}\rangle
               +|\psi_{3}\rangle
               \},                                                                                                            \cr\cr
  |\Theta_{-}\rangle&=& \frac{1}{N_{e}}\{
               \frac{\Omega_{1}}{\varsigma}[\frac{\vartheta^{2}}{2}-(\Omega_{2}^{2}+2\lambda^{2})]|\psi_{1}\rangle
               -\frac{\Omega_{2}}{\varsigma}[\frac{\vartheta^{2}}{2}-(\Omega_{1}^{2}+2\lambda^{2})]|\psi_{5}\rangle          \cr\cr
             &&+\frac{\vartheta({2\lambda^{2}+\varpi})}{2\varsigma}|\phi_{1}\rangle
               +\frac{\vartheta}{2\lambda}|\mu_{2}\rangle
               +|\psi_{3}\rangle
               \},
\end{eqnarray}
with $\varsigma=\lambda(\Omega_{1}^{2}-\Omega_{2}^{2})$,
and $N_{e}$ is the corresponding normalization coefficient.
Whereas the adiabatic condition for the whole system is not always satisfied,
the eigenstates $|\Theta_{+}\rangle$ and
$|\Theta_{-}\rangle$ will be populated and participate
in the evolution of the whole system. On account of a wide disparity between the corresponding eigenvalues of the rest eigenstates and 0, these states can be adiabatically eliminated.
The states $|\Theta_{+}\rangle$ and $|\Theta_{-}\rangle$ are similar to each other,
thus we take $|\Theta_{+}\rangle$ for an example in the following analysis.
The ratio $\tau$ of the coefficients for states $|\phi_{1}\rangle$ and $|\mu_{2}\rangle$ is
\begin{eqnarray}\label{eq 121}
  \tau&=|&{\frac{\vartheta}{\varsigma}[\frac{\vartheta^{2}}{2}-(\frac{\Omega_{1}^{2}+\Omega_{2}^{2}}{2}+2\lambda^{2})]}/
         ({-\frac{\vartheta}{2\lambda}})|                                                                    \cr\cr
      &=&|-\frac{\lambda[\vartheta^{2}-(\Omega_{1}^{2}+\Omega_{2}^{2}+4\lambda^2)]}{\varsigma}|              \cr\cr
      &=&|\frac{2\lambda^{2}+\varpi}{\Omega_{1}^{2}-\Omega_{2}^{2}}|.
\end{eqnarray}
If we set $\Omega_{1}(t)=\zeta\lambda\sin{(\pi/2t_{f})}$ and $\Omega_{2}(t)=\zeta\lambda\cos{(\pi/2t_{f})}$, where $\zeta\lambda$ denotes
the amplitude of the laser pulse,
\begin{eqnarray}\label{eq 122}
  \tau&=&|\frac{2+\sqrt{\zeta^{4}(\sin^{2}{\beta}-\cos^{2}{\beta})^{2}+4}}{\zeta^{2}(\sin^{2}{\beta}-\cos^{2}{\beta})}|.
\end{eqnarray}
It is evident, there is a minimum value and a maximum value of $\tau$, namely,
$\tau_{min}=|(2+\sqrt{\zeta^{4}+4})/\zeta^{2}|$ and $\tau_{max}=\infty$. From the conditions described above, $\zeta<\sqrt{2}$ should be satisfied. When $\zeta=\sqrt{2}$,
$\tau_{min}=1+\sqrt{2}$, and the corresponding ratio of the populations for the states
$|\phi_{1}\rangle$ and $|\mu_{2}\rangle$ is $\tau^{2}=3+2\sqrt{2}$. The result reveals that, with the limits to the parameters of $\tau_{min}$,
the population of the state $|\mu_{2}\rangle$ is still much less than that of the
state $|\phi_{1}\rangle$. And the population of the state $|\phi_{1}\rangle$
keeps in a small value during the evolution of the whole system (this will be
analyzed in detail later). Afterwards, we deduce that the population for the
state $|\mu_{2}\rangle$ can be neglected all the time during the evolution.
From Eq. (\ref{eq 117}), we find that the state $|\psi_{3}\rangle$ can only
be transformed from the state $|\mu_{2}\rangle$. Since the
population for the state $|\mu_{2}\rangle$ is neglected all the time,
$|\psi_{3}\rangle$ is considered as an independent state of the whole
system. That is, the whole system is regarded as a three-level
single-atom system even when the Zeno condition is not well met.
However, as the result of strong coupling between the states $|\mu_{2}\rangle$ and
$|\psi_{3}\rangle$, very little population for the state
$|\mu_{2}\rangle$ can lead to a rapid increase in the
population for the state $|\psi_{3}\rangle$. The effects of the subsystems $S_{1}$ and $S_{3}$ in the population
transfer of the whole system are embodied by the dark state $|Dark\rangle$.
And the intermediate state $|\psi_{3}\rangle$ will become the key point of the combined effect of
the subsystems $S_{1}$ and $S_{3}$ for assisting the population transfer.
The population of the intermediate state $|\psi_{3}\rangle$ is mainly dominated by the ratio
$r=\Omega_{1}(t)\Omega_{2}(t)/(N_{2}\lambda)$ according to Eq. (\ref{eq 120}).
For simplicity, we set $t=t_{f}/2$
(the population of the state $|\psi_{3}\rangle$ is the maximum when $t=t_{f}/2$) such that
\begin{eqnarray}\label{eq119}
  r=\frac{\pi\cot{\epsilon}}{\sqrt{(2\sqrt{2}\lambda t_{f})^{2}+(\pi\cot{\epsilon})^2}},
\end{eqnarray}
i.e., when $\epsilon$ is a constant value, the larger the
interaction time $\lambda t_{f}$ is, the less the population of the
intermediate state $|\psi_{3}\rangle$ is. From Refs.
\cite{XCILARDGJGMPrl10,XCJGMPra12,MLLTSYXJS13}, the
essence of FPT in the invariant-based inverse
engineering is increasing the populations of some intermediate
states under certain conditions. Now, if we suitably increase the
population of the intermediate state $|\psi_{3}\rangle$ (actually,
the population of the state $|\psi_{3}\rangle$ is increased by very
slightly increasing the population of $|\mu_{2}\rangle$) with very
slightly destroying the conditions for the perfect FPT in the main
subsystem $S_{2}$, the transfer will be much faster for the
relation between the population of the state $|\psi_{3}\rangle$ and
the interaction time is inversely proportional when $\epsilon$ is a
constant value.

\begin{figure}
 \renewcommand\figurename{\small FIG.}
 \centering \vspace*{8pt} \setlength{\baselineskip}{10pt}
 \subfigure[]{
 \includegraphics[scale = 0.33]{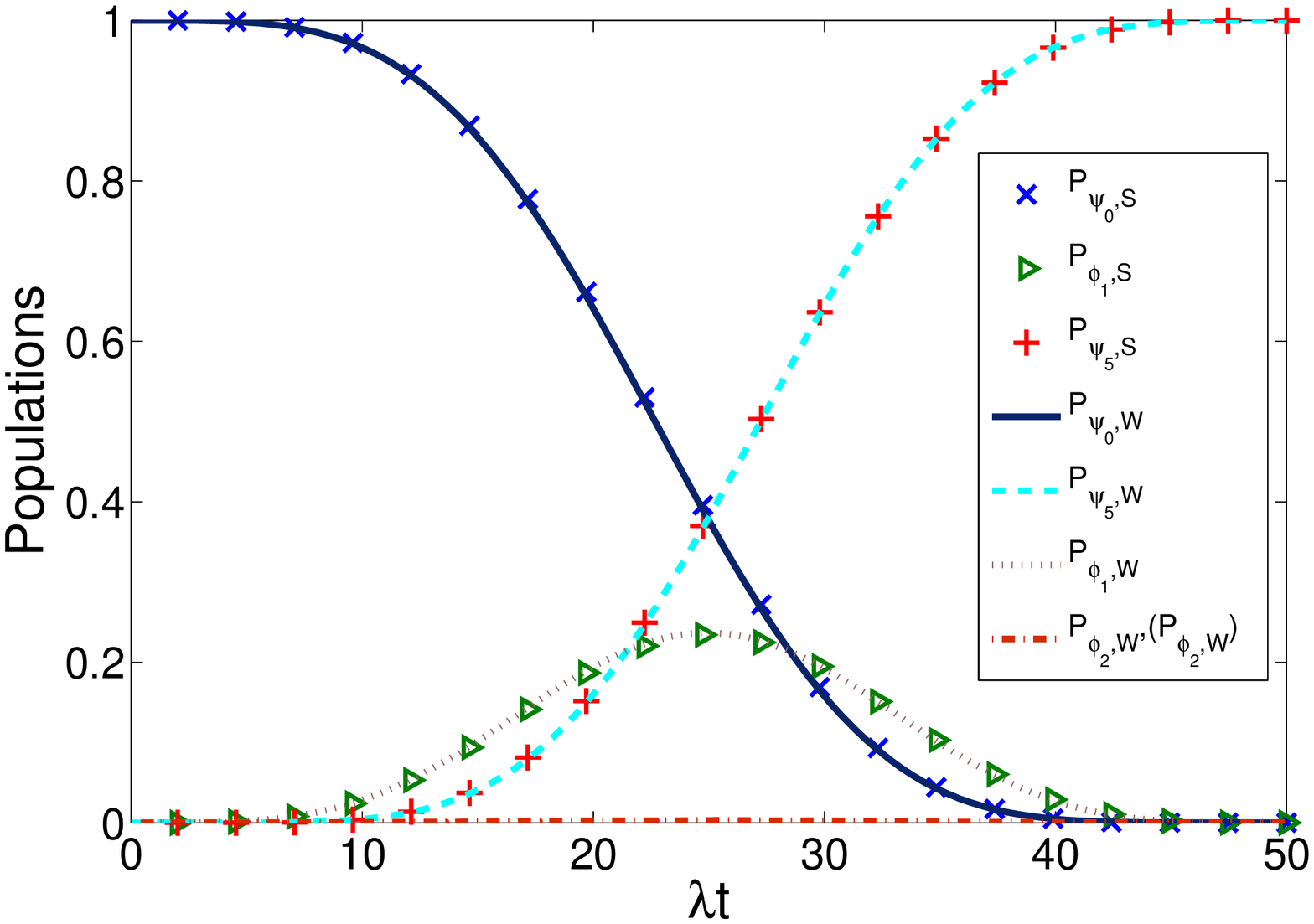}}
 \subfigure[]{
 \includegraphics[scale = 0.33]{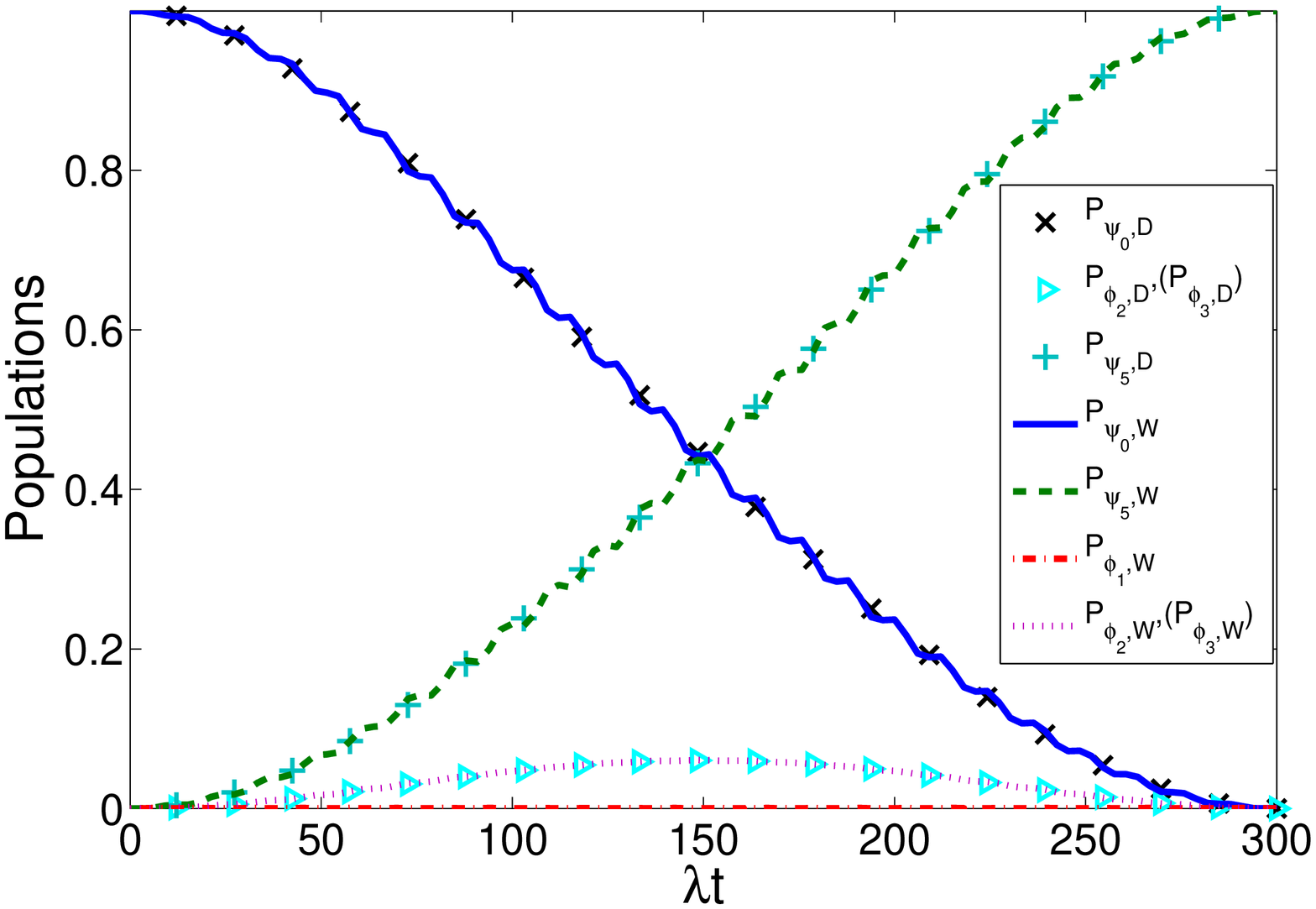}}
 \caption{
    (a) The comparison between the populations transfer governed by the total Hamiltonian $H_{I}$ and that governed by the Hamiltonian
        $H_{S_{2}}$ when $\lambda t_{f}=50$ and $\epsilon=\arcsin{1/4}$.
    (b) The comparison between the populations transfer governed by the total Hamiltonian $H_{I}$ and that governed by the dark state $|Dark\rangle$
        when $\lambda t_{f}=300$ and $\epsilon=\arcsin{1/100}$.
          }
 \label{HiHs2Dark}
\end{figure}

The validity of the above theoretical
analysis will be numerically proved in the following. First, the population
transfer of the whole system is an ideal FPT when the Zeno condition
is greatly satisfied. Figure. \ref{HiHs2Dark} (a) shows the
comparison between the population transfer governed by the total
Hamiltonian $H_{I}$ according to Eq. (\ref{eq1}) and that governed
by the Hamiltonian of subsystem $S_{2}$ according to Eq. (\ref{eq7})
when $\lambda t_{f}=50$ and $\epsilon=\arcsin{0.25}$ [the Zeno
condition $\lambda\gg\Omega_{k}(t)$ can be satisfied very well],
where the markers with different styles and colors represent the
time evolution of the populations governed by the subsystem
Hamiltonian $H_{S_{2}}$ for the states $|\psi_{0}\rangle$,
$|\psi_{5}\rangle$, and $|\phi_{1}\rangle$, respectively, and the
curves with different styles and colors represent the time evolution
of the populations governed by the total Hamiltonian $H_{I}$ for the
states $|\psi_{0}\rangle$, $|\psi_{5}\rangle$, $|\phi_{1}\rangle$,
and $|\phi_{2}\rangle$ ($|\phi_{3}\rangle$), respectively, and the
superscripts $S$ and $W$ represent the Hamiltonian of the subsystem
$S_{2}$ and the total Hamiltonian $H_{I}$, respectively. The
population for a state $|\psi\rangle$ is given through the relation
$P=|\langle\psi|\rho(t)|\psi\rangle|$, where $\rho(t)$ is the
density operator of the system at any time $t$. All the time, the
populations of the states $|\phi_{2}\rangle$ and $|\phi_{3}\rangle$
remain negligible, and the time evolution of the system
governed by the total Hamiltonian $H_{I}$ is exactly the same with
the time evolution of the system governed by the subsystem
Hamiltonian $H_{S_{2}}$ if we neglect the states $|\phi_{2}\rangle$
and $|\phi_{3}\rangle$, that is, we quote Chen \emph{et al.} as saying that
the whole system evolves along the dark state $|\Psi_{0}(t)\rangle$
and an FPT of the whole system can be perfectly achieved.
In fact, the dark state $|\Psi_{0}(t)\rangle$ can't
faultlessly explain the evolution of the system; the system evolves
along a special way which is very similar to a dark state,
and we name it ``dark-like state'' for short.
This special state has the form $|D_{like}(t)\rangle=\frac{1}{N_{like}}
[\alpha_{1}(t)|\psi_{1}\rangle+\alpha_{2}(t)|\psi_{5}\rangle
+\alpha_{3}|\phi_{1}\rangle+\alpha_{4}(t)|\mu_{1}\rangle]$.
In the present case, as the Zeno condition is satisfied,
the state $|\mu_{1}\rangle$ is negligible and the ``dark-like state'' can be
simplified as $|D'_{like}(t)\rangle=\frac{1}{N'_{like}}
[\alpha_{1}(t)|\psi_{1}\rangle+\alpha_{2}(t)|\psi_{5}\rangle
+\alpha_{3}(t)|\phi_{1}\rangle]$.

We confirm that the evolution of the whole system is completely governed
by the dark state $|Dark\rangle$ with completely destroying the
conditions for the FPT in the subsystem $S_{2}$ [when $\sin{\gamma}$
is very close to zero, the invariant Hermitian operator $I_{S_{2}}$
according to Eq. (\ref{eq9}) equals to the Hamiltonian $H_{S_{2}}$,
and the system is just an ordinary system based on STIRAP]. Figure
\ref{HiHs2Dark} (b) shows the comparison between the population
transfer governed by the total Hamiltonian $H_{I}$ according to Eq.
(\ref{eq1}) and that governed by the dark state $|Dark\rangle$
according to Eq. (\ref{eq 118}) when $\lambda t_{f}=300$ and
$\epsilon=\arcsin{1/100}$ (the condition for STIRAP can be
satisfied), where the markers with different styles and colors
represent the time evolution of the populations governed by the dark
state $|Dark\rangle$ for the states $|\psi_{0}\rangle$,
$|\phi_{2}\rangle$ ($|\phi_{3}\rangle$), and $|\psi_{5}\rangle$,
respectively, and the curves with different styles and colors
represent the time evolution of the populations governed by the
total Hamiltonian $H_{I}$ for the states $|\psi_{0}\rangle$,
$|\psi_{5}\rangle$, $|\phi_{1}\rangle$, and $|\phi_{2}\rangle$
($|\phi_{3}\rangle$), respectively, and superscript $D$ represents
the dark state $|Dark\rangle$. Similar to Fig. \ref{HiHs2Dark} (a),
the time evolution of the whole system is almost
absolutely governed by the dark state $|Dark\rangle$ when the dark
state $|\Psi_{0}(t)\rangle$ is inoperative for the evolution of the
whole system (the conditions for the FPT in the subsystem $S_{2}$
are completely ungratified). Contrast Fig. \ref{HiHs2Dark} (a) with
Fig. \ref{HiHs2Dark} (b); the interaction time needed
for the FPT in the invariant-based inverse engineering
satisfying the Zeno condition is much shorter than the
population transfer in an ordinary STIRAP, that is, we have speeded up the population transfer of ground states in a two-atom system with
a composed system including the QZD and the invariant-based inverse engineering.

\begin{figure}
 \renewcommand\figurename{\small FIG.}
 \centering \vspace*{8pt} \setlength{\baselineskip}{10pt}
 \subfigure[]{
 \includegraphics[scale = 0.33]{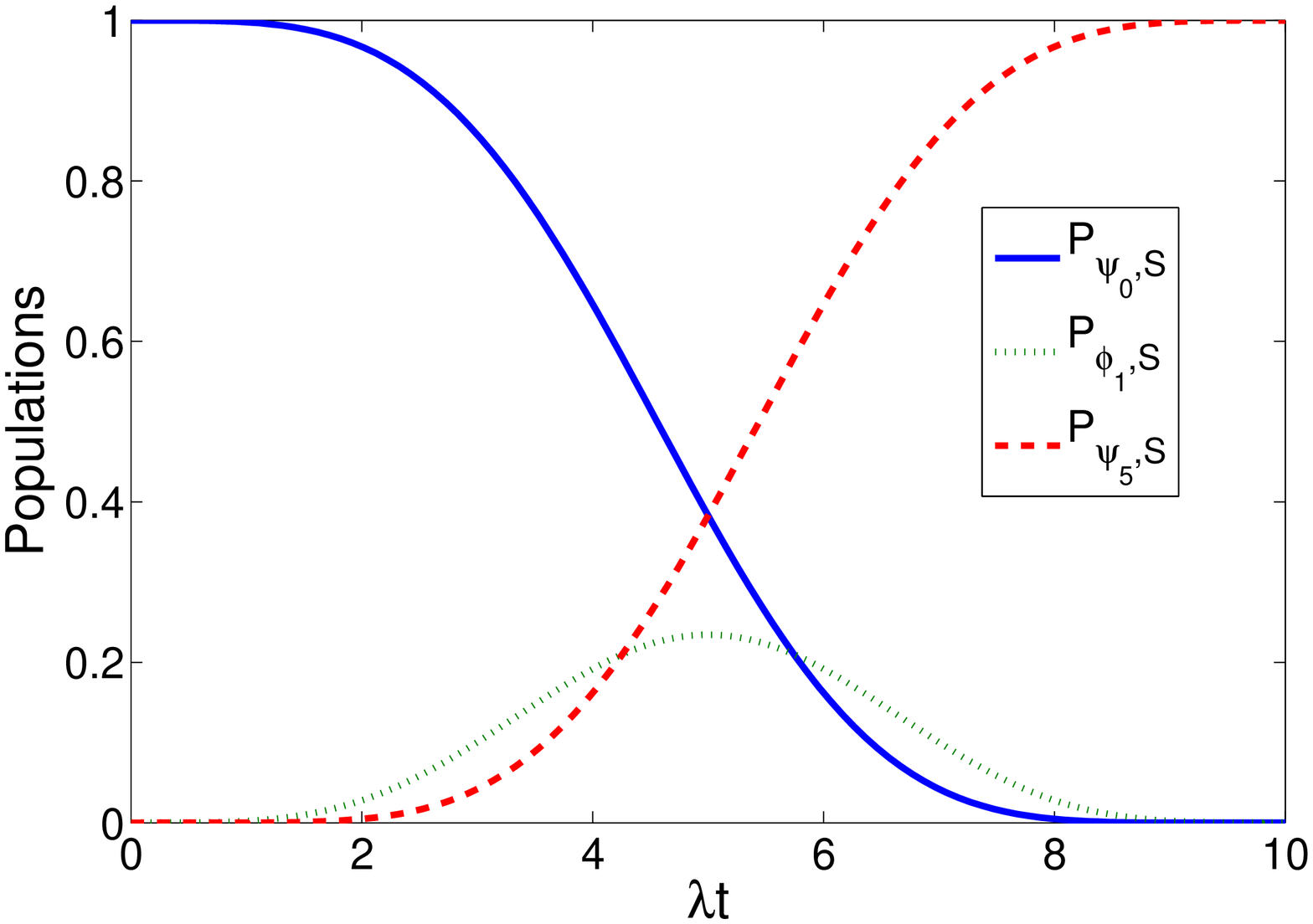}}
 \subfigure[]{
 \includegraphics[scale = 0.33]{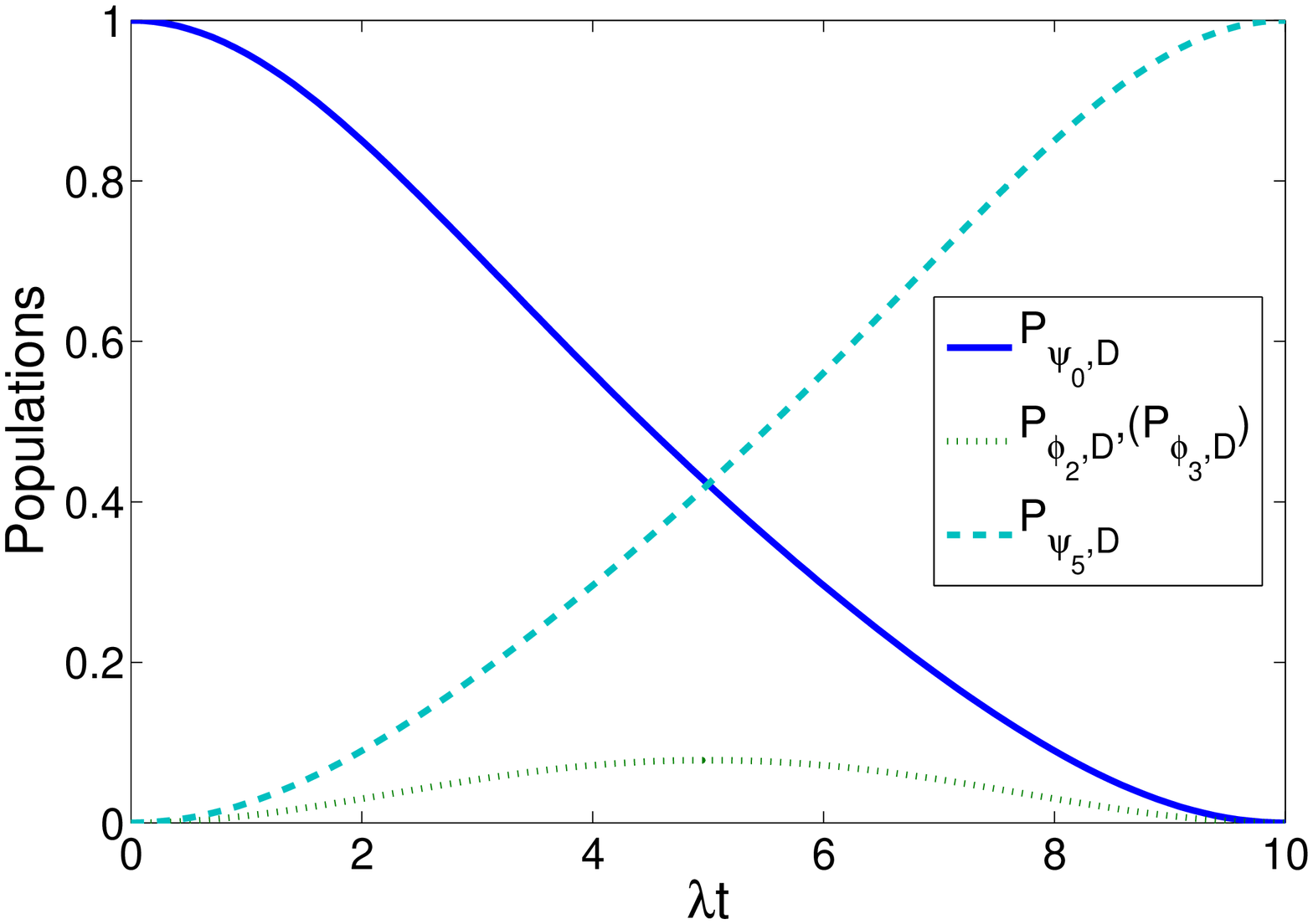}}
 \subfigure[]{
 \includegraphics[scale = 0.33]{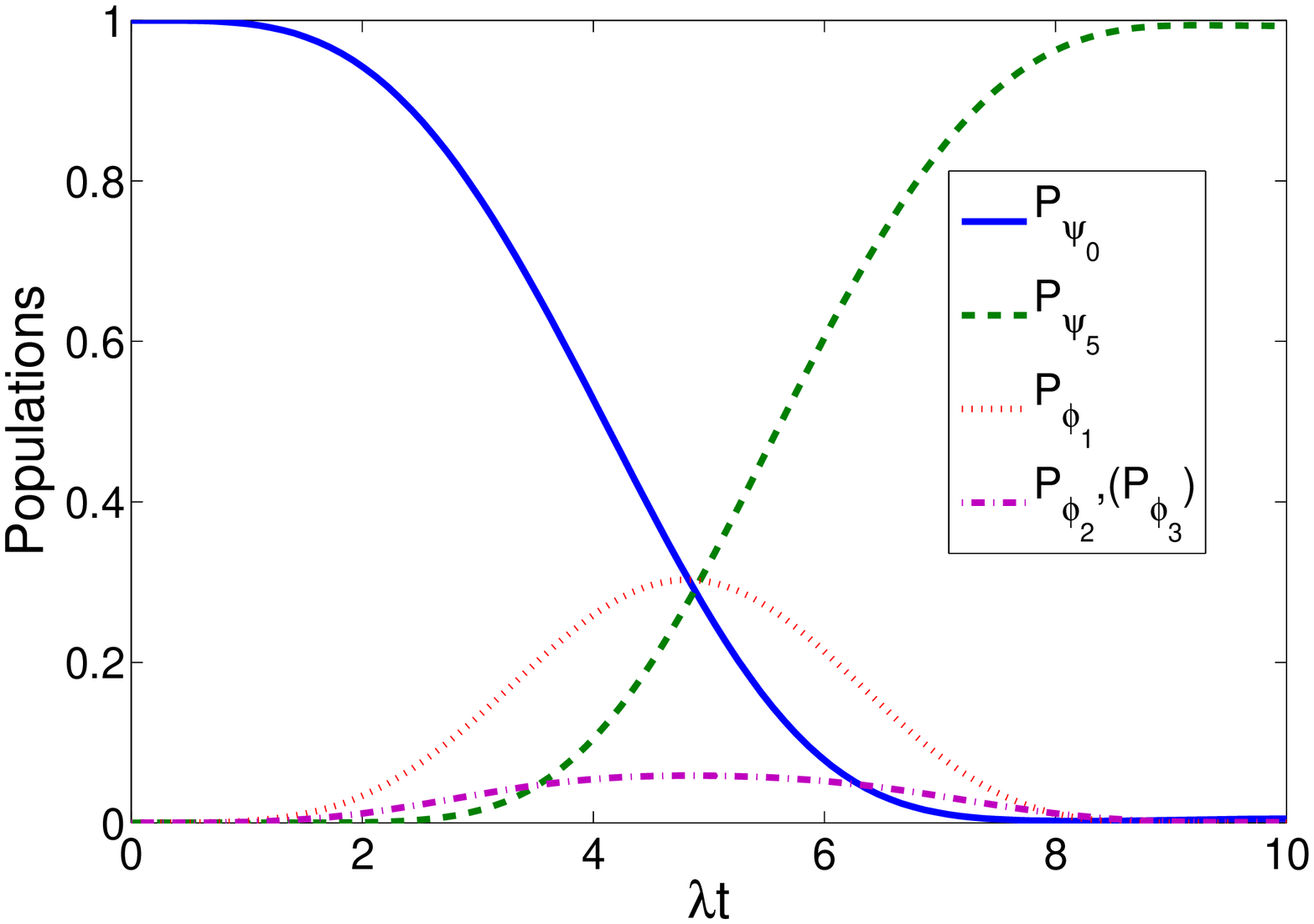}}
 \subfigure[]{
 \includegraphics[scale = 0.33]{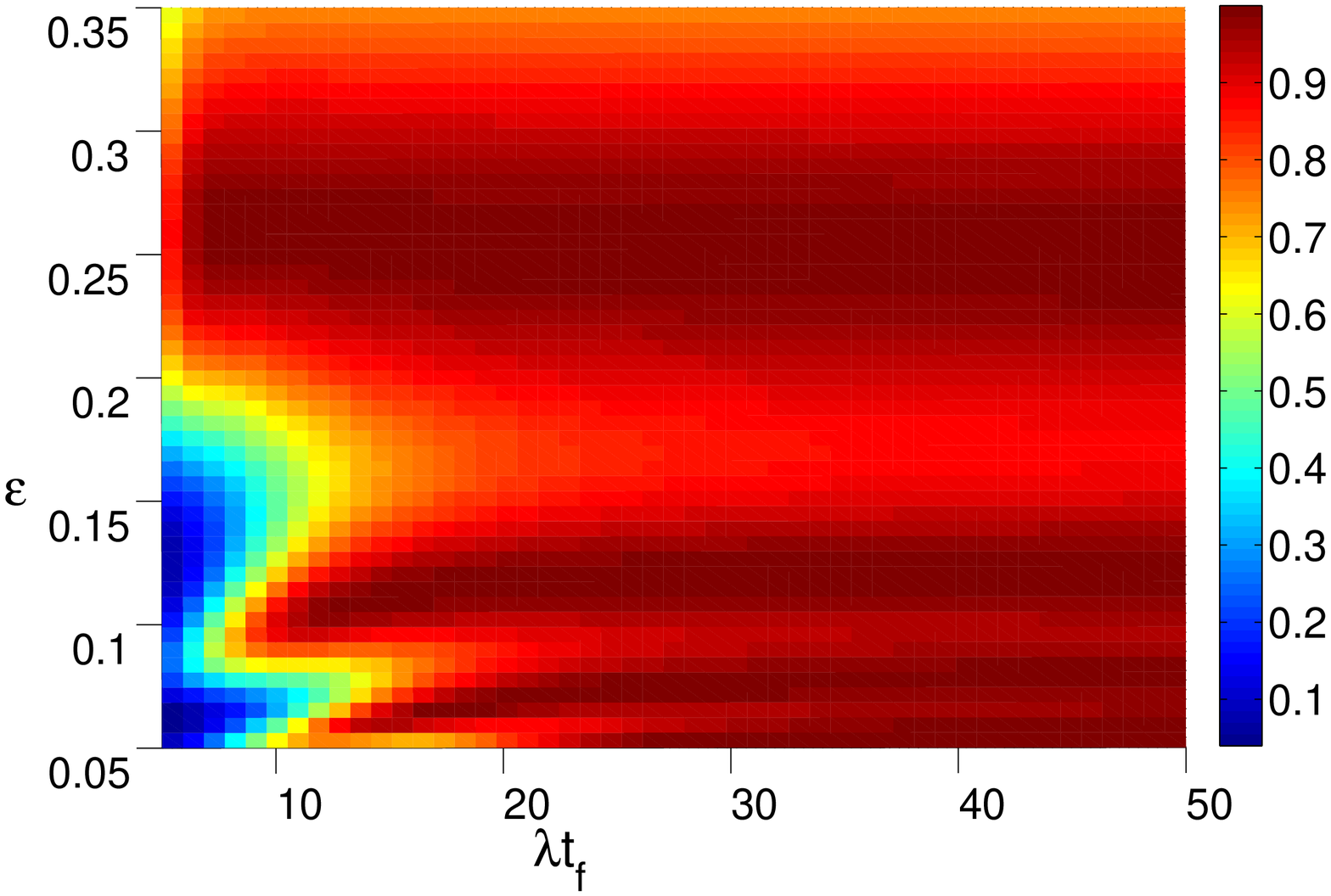}}
 \caption{
    (a) The time evolution of populations governed by the Hamiltonian $H_{S_{2}}$ for the states $|\psi_0\rangle$, $|\phi_1\rangle$, and $|\psi_{5}\rangle$
        when $\lambda t_{f}=10$ and $\epsilon=\arcsin{0.25}$.
    (b) The time evolution of populations governed by the dark $|Dark\rangle$ for the states $|\psi_{0}\rangle$, $|\phi_{2}\rangle$ ($|\phi_{3}$), and
        $|\psi_{5}\rangle$ when $\lambda t_{f}=10$ and $\epsilon=\arcsin{0.25}$.
    (c) The time evolution of populations governed by the total Hamiltonian $H_{I}$ for the states $|\psi_{0}\rangle$, $|\psi_{5}\rangle$, $|\phi_{1}\rangle$,
        and $|\phi_{2}\rangle$ ($|\phi_{3}$) when $\lambda t_{f}=10$ and $\epsilon=\arcsin{0.25}$.
    (d) The fidelity $F$ of the target state $|\psi_{5}\rangle$ versus the value of $\epsilon$ and the interaction time $\lambda t_{f}$.
          }
 \label{HIHs2DarkN1}
\end{figure}

\begin{figure}
 \renewcommand\figurename{\small FIG.}
 \centering \vspace*{8pt} \setlength{\baselineskip}{10pt}
 \subfigure[]{
 \includegraphics[scale = 0.33]{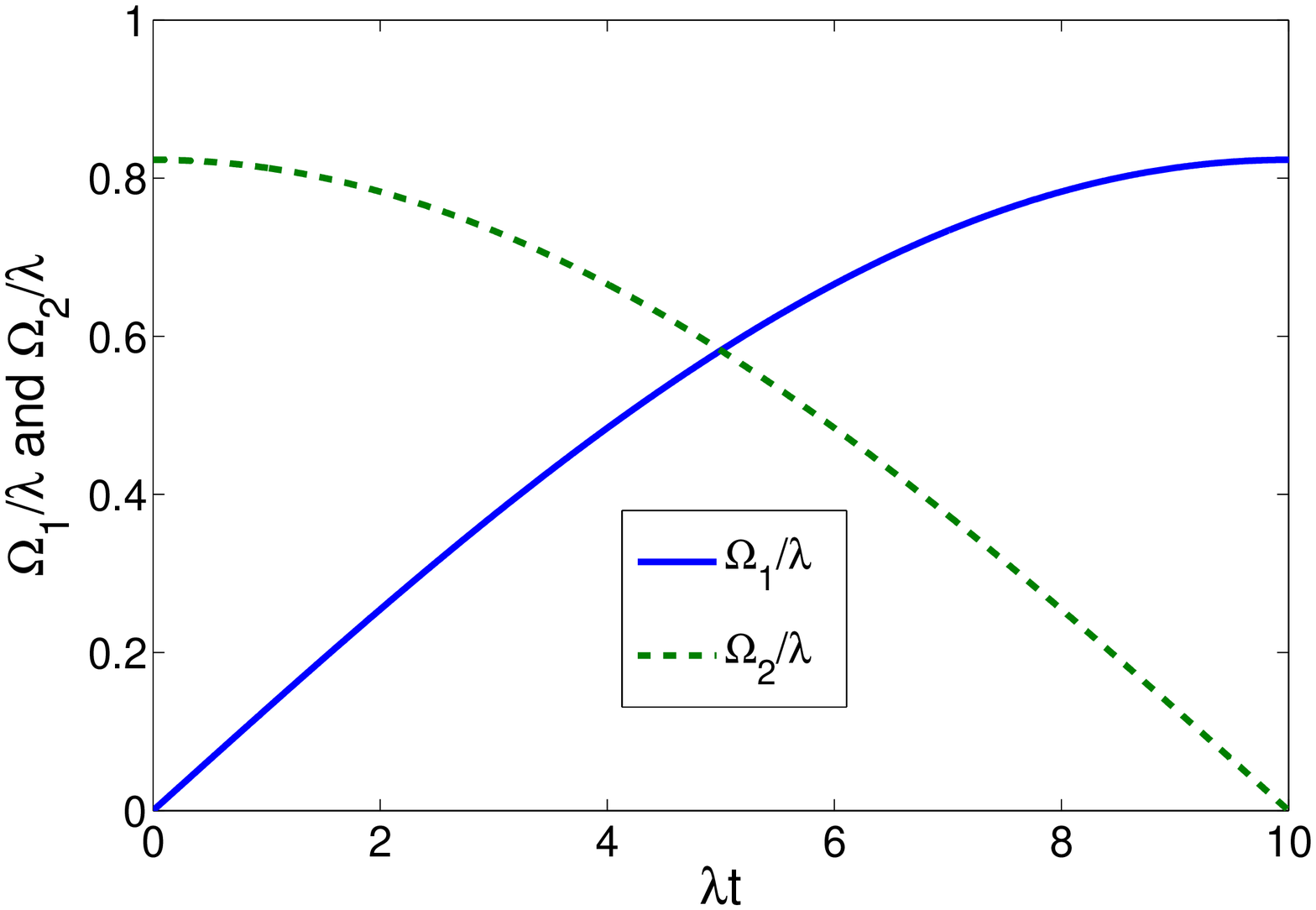}}
 \subfigure[]{
 \includegraphics[scale = 0.33]{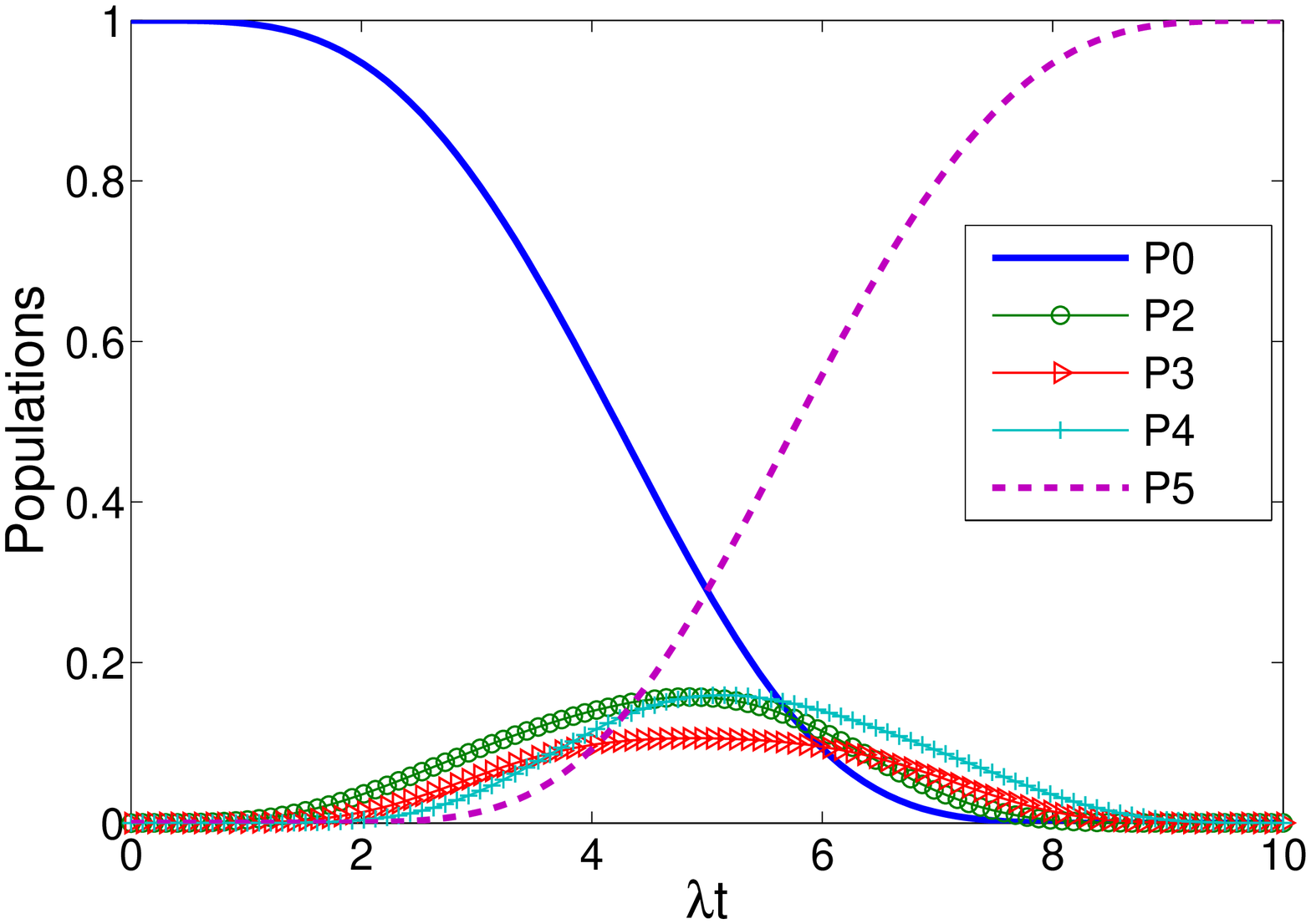}}
 \caption{
    (a) The time dependence of the laser fields $\Omega_1(t)$ and $\Omega_2(t)$ when $\epsilon=0.2636$ and $\lambda t_f=10$.
    (b) Time evolution of the populations for the states $|\psi_0\rangle$, $|\psi_2\rangle$,
    $|\psi_3\rangle$, $|\psi_4\rangle$, and $|\psi_5\rangle$ when $\epsilon=0.2636$ and $\lambda t_f=10$.
          }
 \label{O1O2P12345}
\end{figure}

\begin{figure}
 \renewcommand\figurename{\small FIG.}
 \centering \vspace*{8pt} \setlength{\baselineskip}{10pt}
 \subfigure[]{
 \includegraphics[scale = 0.33]{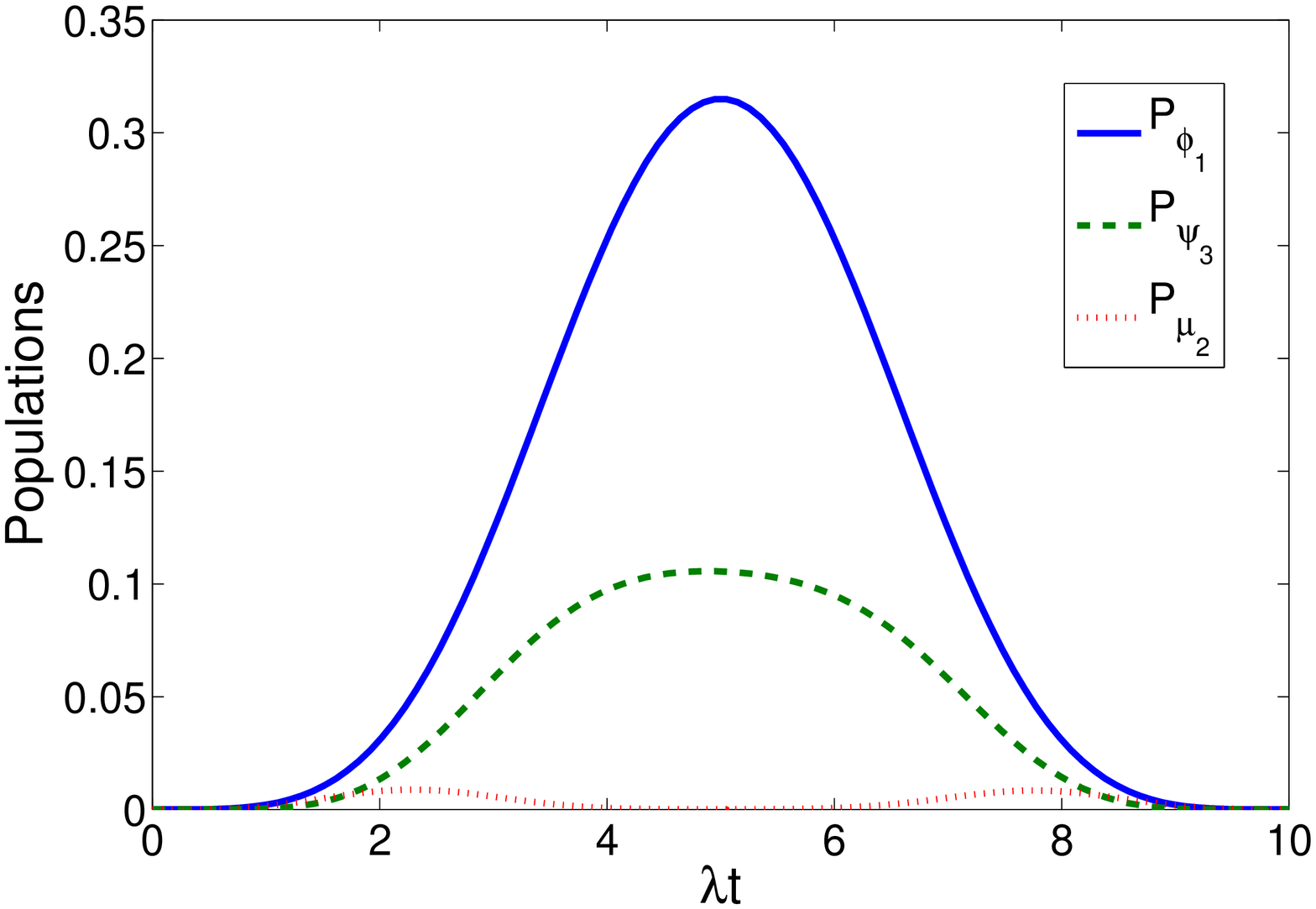}}
 \subfigure[]{
 \includegraphics[scale = 0.33]{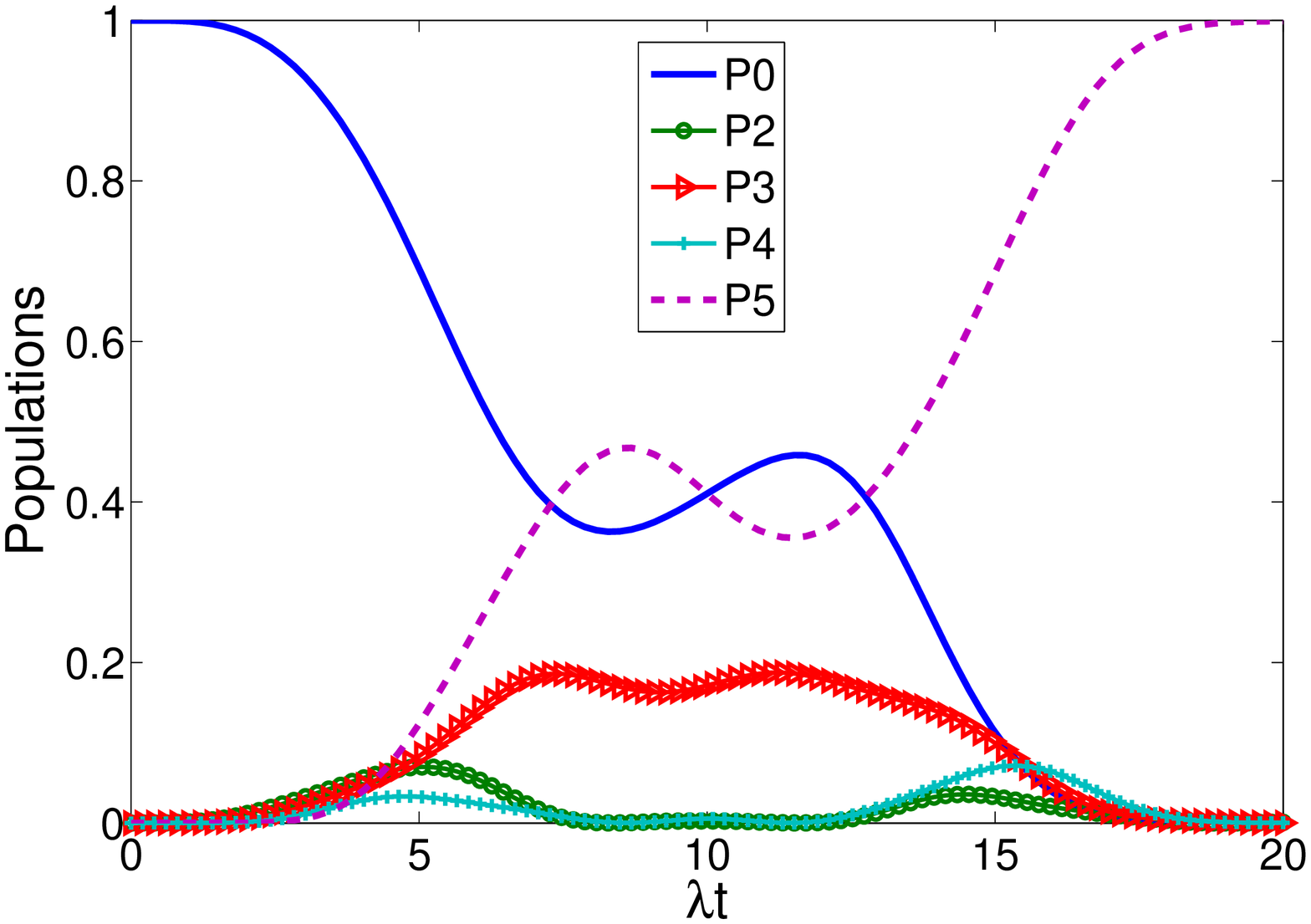}}
 \caption{
    (a) Time evolution of the populations for the  intermediate states $|\phi_{1}\rangle$,
        $|\psi_{3}\rangle$, and $|\mu_{2}\rangle$ when $\epsilon=0.2636$ and $\lambda t_f=10$.
    (b) Time evolution of the populations for the states $|\psi_0\rangle$, $|\psi_2\rangle$,
        $|\psi_3\rangle$, $|\psi_4\rangle$, and $|\psi_5\rangle$ when $\epsilon=0.1196$ and $\lambda t_f=20$.
          }
 \label{P234N2}
\end{figure}

Moreover, we further shorten the interaction time by combining
the effect of the subsystem $S_{2}$ with the effect of the dark
state $|Dark\rangle$ (the combined effect of the subsystems $S_{1}$
and $S_{3}$). Figure \ref{HIHs2DarkN1} (a) shows the time evolution of
the populations governed by the Hamiltonian $H_{S_{2}}$ for the
states $|\psi_{0}\rangle$, $|\phi_{1}\rangle$, and
$|\psi_{5}\rangle$, Fig. \ref{HIHs2DarkN1} (b) shows the time
evolution of the populations governed by the dark state
$|Dark\rangle$ for the states $|\psi_{0}\rangle$, $|\phi_{2}\rangle$
($|\phi_3\rangle$), and $|\psi_{5}\rangle$, and Fig.
\ref{HIHs2DarkN1} (c) shows the time evolution of the populations
governed by the total Hamiltonian $H_{I}$ for the states
$|\psi_{0}\rangle$, $|\phi_{1}\rangle$, $|\phi_{2}\rangle$
($|\phi_3\rangle$), and $|\psi_{5}\rangle$. Figs. \ref{HIHs2DarkN1}(a)-\ref{HIHs2DarkN1}(c)
are plotted with $\epsilon=\arcsin{0.25}$ ($N=1$)
and $\lambda t_{f}=10$. Contrast Fig. \ref{HIHs2DarkN1} (c) with
Figs. \ref{HIHs2DarkN1} (a) and \ref{HIHs2DarkN1} (b); the time evolution of the whole system governed by the
combined effect of the subsystem $S_{2}$ and the dark state
$|Dark\rangle$ is a little more complex than that governed by the
effect of directly adding these two effects together. It can be seen
from Fig. \ref{HIHs2DarkN1} (c) that the population of the
target state $|\psi_{5}\rangle$ is only $99.35\%$ when $t=t_{f}$.
The reason for these results can be understood by the conditions
(the Zeno condition, the condition for STIRAP, etc.) for whether an
ideal FPT governed by the subsystem $S_{2}$ or an ideal population
transfer governed by the dark state $|Dark\rangle$ can not be
satisfied very well, actually the population transfer from the initial
state to the target state along a dark-like state
$|Dark\rangle_{like}$ which will be discussed in detail elsewhere.
Due to the slightly
 populated intermediate state $|\mu_{2}\rangle$, the whole system can not be faultlessly considered as a
three-level single-atom system, and
the optimal value of $\epsilon$ for the whole system will not
faultlessly satisfy the condition $(\sin{\epsilon})^{-1}=4N$
($N=1,2,3,\cdots$). Reselecting the optimal value
of $\epsilon$ becomes a necessity. We plot the fidelity $F$ of the target state
$|\psi_{5}\rangle$ versus the value of $\epsilon$ and the
interaction time $\lambda t_{f}$ in FIG. \ref{HIHs2DarkN1} (d). The
fidelity $F$ for the target state $|\psi_{5}\rangle$ is given
through the relation $F=|\langle\psi_{5}|\rho(t_{f})|\psi_{5}\rangle|$, where
$\rho(t_{f})$ is the density operator of the system at the time
$t_{f}$ by solving the differential equation
$\dot{\rho}=i[\rho,H_{I}]$. When $\lambda
t_{f}=10$, the optimal value of $\epsilon$ for the highest fidelity
($F=1$) of the state $|\psi_{5}\rangle$ is about $0.2636$,
meanwhile, the minimum value of $\lambda t_{f}$ is
only about $7.3$ for a perfect FPT ($F=1$ for the target state when
$t=t_{f}$), even when $\lambda t_{f}=6.4$ and $\epsilon\approx0.26$,
the fidelity of the target state is higher than $99\%$ (when
$\lambda t_{f}<6$, the whole system can not be considered as a
three-level single-atom system since the state $|\mu_{2}\rangle$ is
populated too much and can not be neglected). What is more, this method is insensitive to the fluctuations of
$\epsilon$ and the interaction time $\lambda t_{f}$, and is also insensitive to the amplitude of the laser pulses and the coupling
constant $\lambda$. For convenient discussion, we suitably choose three sets of
parameters $\{\epsilon=0.2636,\ \lambda t_{f}=10\}$,
$\{\epsilon=0.1196,\ \lambda t_{f}=20\}$, and $\{\epsilon=0.0810,\
\lambda t_{f}=40\}$, corresponding $N=1$, $N=2$, and $N=3$,
respectively. Figure \ref{O1O2P12345} (a) shows the time dependence of
the Rabi frequencies for the atoms when $\epsilon=0.2636$ and
$\lambda t_{f}=10$. The ratio
$\Omega_{k}^{max}/\lambda$ (here the superscript $max$ denotes the
maximum value of $\Omega_{k}$) is $0.8232$ which meets the
conditions mentioned above. And Fig. \ref{O1O2P12345} (b) shows the
time evolution of the populations for states $|\psi_0\rangle$,
$|\psi_{2}\rangle$, $|\psi_{3}\rangle$, $|\psi_{4}\rangle$, and
$|\psi_{5}\rangle$. After reselecting the
optimal value of $\epsilon$, a perfect population transfer from the
initial state $|\psi_{0}\rangle$ to the target state
$|\psi_{5}\rangle$ (the population of the target state
$|\psi_{5}\rangle$ is $1$ when $t=t_{f}$) can be achieved. Figure
\ref{P234N2} (a) shows the time evolution of the populations for the
states $|\phi_{1}\rangle$ ($|\mu_{1}\rangle$), $|\psi_{3}\rangle$,
and $|\mu_{2}\rangle$. The population of
$|\mu_{2}\rangle$ remain negligible all the time even with
$\zeta=0.8232$. Actually, Fig. \ref{P234N2} (a) explains the essence
of FPT. The intermediate states $|\phi_{1}\rangle$,
$|\psi_{3}\rangle$, and $|\mu_{2}\rangle$ are usually neglected in
the schemes in the view of STIRAP and QZD. However, these states are
necessary for the transfer from the initial state $|\psi_{0}\rangle$ to
the target state $|\psi_{5}\rangle$. They link the whole system
together just like brittle strings; the evolution of the system is
interdictory without the participation of these intermediate states. 
By increasing the populations of intermediate states in a
certain period of time, just like broadening the channels for the
transition between $|\psi_{0}\rangle\leftrightarrow|\psi_{5}\rangle$
in a certain period of time, the transition could be much
faster. Figures. \ref{O1O2P12345} and \ref{P234N2} (a) are plotted
when $\lambda t_{f}=10$ and $\epsilon=0.2636$. Figure \ref{P234N2} (b)
shows the populations for the states $|\psi_{0}\rangle$,
$|\psi_{2}\rangle$, $|\psi_{3}\rangle$, $|\psi_{4}\rangle$, and
$|\psi_{5}\rangle$ when $\lambda t_{f}=20$ and $\epsilon=0.1196$.
Contrast Fig. \ref{O1O2P12345} (b) with Fig. \ref{P234N2} (b); 
it turns out that a longer interaction time is required, i.e.,
$t_f=20/\lambda$, when $\epsilon=0.1196$ for achieving the target
state, and the population of $|\psi_3\rangle$ only changes a little
while the populations of $|\psi_{2}\rangle$ and $|\psi_{4}\rangle$
change a lot. The reason for needing a longer interaction time is that a
smaller $\epsilon$ causes larger amplitudes of the laser pulses, and
a relatively larger $\lambda t_{f}$ should be chosen to satisfy the
conditions above. As narrated above, the
population of $|\psi_{3}\rangle$ is only governed by the combined
effect of subsystems $S_{1}$ and $S_{3}$. The ratio $r$ governs the population of the state
$|\psi_{3}\rangle$ according to eq. (\ref{eq119}). We obviously have to have the ratio $r=0.4195$ when
$\lambda t_{f}=20$ and $\epsilon=0.1196$, and $r=0.3806$ when
$\lambda t_{f}=10$ and $\epsilon=0.2636$. This immediately implies, by varying $\lambda t_{f}$ and $\epsilon$ at a similar rate, the corresponding ratio $r$ shifts only a little bit. That is the reason why the population of
$|\psi_{3}\rangle$ almost keeps unchanging when $\lambda t_{f}$ and
$\cot{\epsilon}$ are changing similarly.

In particular, we contrast this method with an ordinary method based on QZD with the similar model.
When the Zeno condition $\lambda\gg\Omega_{k}$ is satisfied
and the laser pulses are independent of time, based on the QZD,
an effective Hamiltonian of the system is
\begin{eqnarray}\label{eq223}
  H_{eff}=\frac{\Omega_{1}}{\sqrt{2}}|\psi_{0}\rangle\langle\phi_{1}|+
          \frac{\Omega_{2}}{\sqrt{2}}|\psi_{5}\rangle\langle\phi_{1}|+H.c.,
\end{eqnarray}
and the general evolution form of eq. (\ref{eq223}) at time $t$ is
\begin{eqnarray}\label{eq224}
  |\psi(t)\rangle=\frac{1}{2\chi^{2}}(\Omega_{1}^{2}\cos{\chi t}+\Omega_{2}^{2})|\psi_{0}\rangle
                   -i\sin{\chi t}|\phi_{1}\rangle
                   +\frac{1}{2\chi^{2}}(\Omega_{1}\Omega_{2}\cos{\chi t}-\Omega_{1}\Omega_{2})|\psi_{5}\rangle,
\end{eqnarray}
with $\chi=\sqrt{(\Omega_{1}^{2}+\Omega_{2}^{2})/2}$. When we
choose $t=t_{f}=\pi/\chi$ and $\Omega_{Z}=\Omega_{1}=\Omega_{2}$, the
target state $|\psi_{5}\rangle$ is obtained. For $\lambda t_{f}=\pi/\Omega_{Z}$, if $\Omega_{Z}=0.1\lambda$ (almost the limitation of the value
of $\Omega_{Z}$ for satisfying the Zeno condition), $t_{f}\simeq31.416\lambda$.
The maximal population of the intermediate $|\phi_1\rangle$
during the evolution of the whole system is $50\%$ when $t=0.5t_{f}$,
that means the influence of decoherence caused by the spontaneous emission is
very great. The minimum effective interaction time $\lambda t_{f}$ as mentioned above, however, is only about 7.2. As noted earlier, the QZD is sensitive to variations in some parameters, especially
the interaction time.
Whereas, this method is insensitive to variations in most of the parameters.
Compare to the method based on STRIRAP and QZD, this method has superiority to some extent.

\begin{figure}
 \renewcommand\figurename{\small FIG.}
 \centering \vspace*{8pt} \setlength{\baselineskip}{10pt}
 \subfigure[]{
 \includegraphics[scale = 0.33]{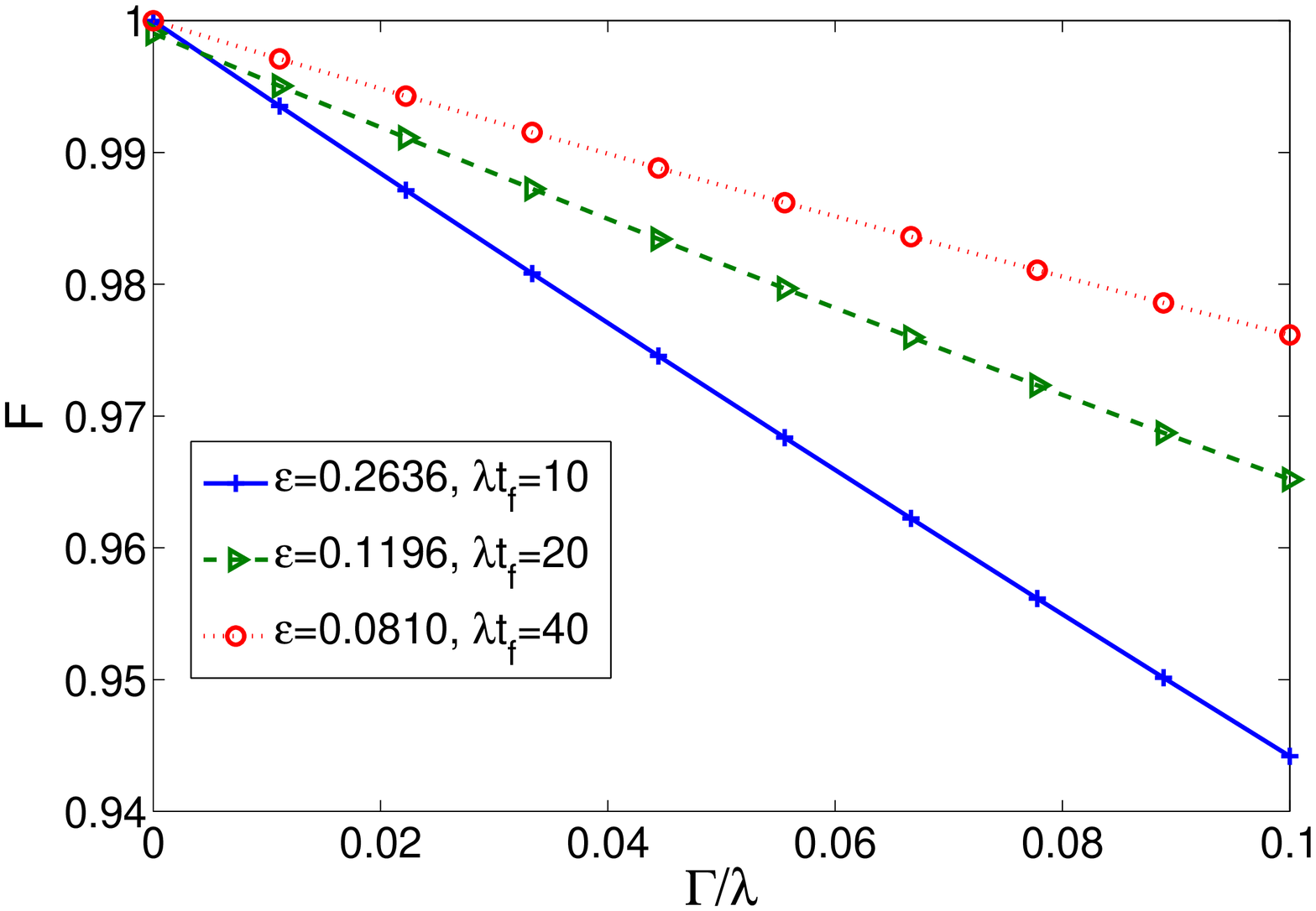}}
 \subfigure[]{
 \includegraphics[scale = 0.33]{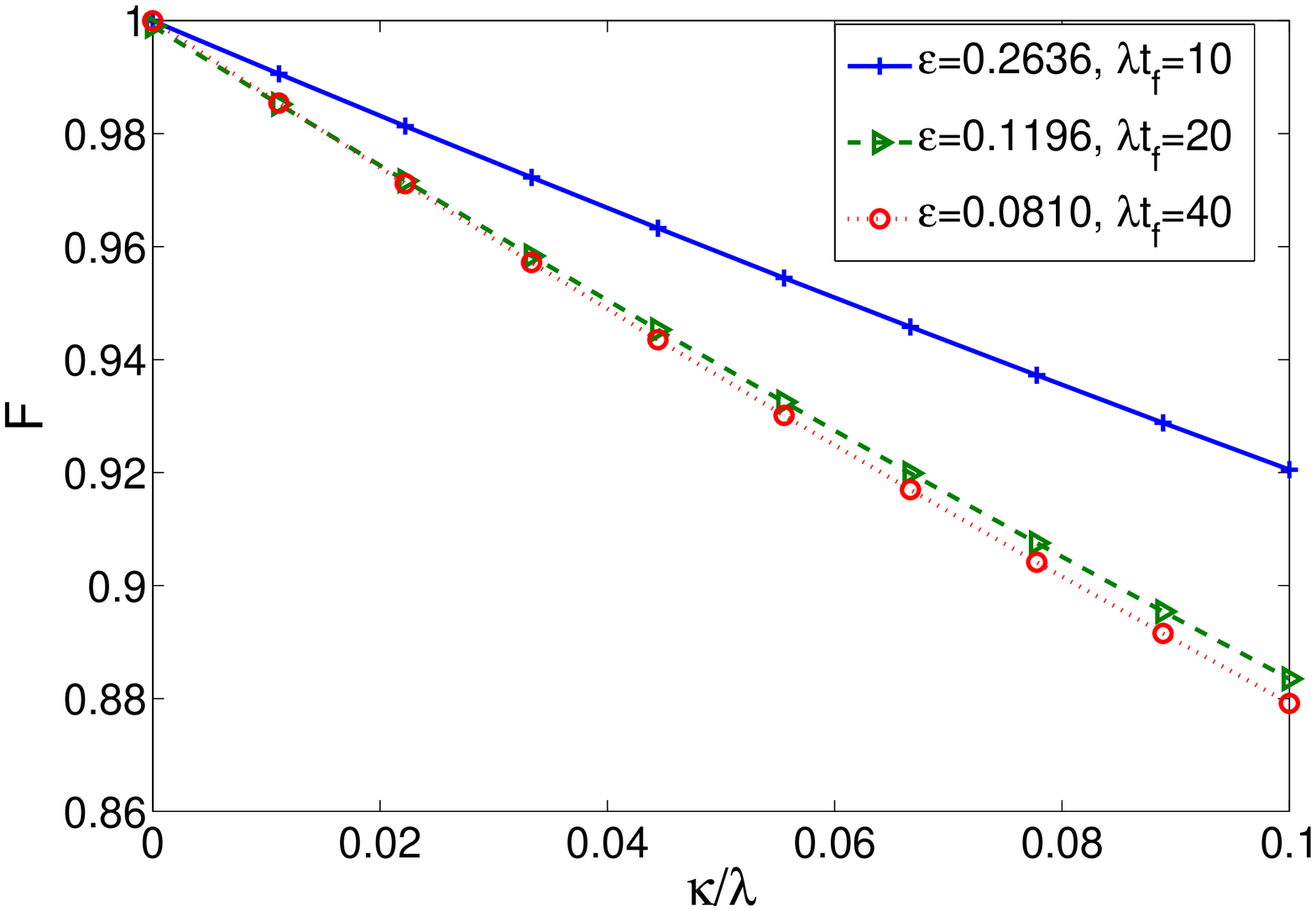}}
 \caption{
    (a) The influence of spontaneous emission $\Gamma/\lambda$ on the fidelity $F$ of the target
        state $|\psi_5\rangle$ under different conditions when the decay of cavity $\kappa=0$.
    (b) The influence of decay of cavity $\kappa/\lambda$ on the fidelity $F$ of the target
        state $|\psi_5\rangle$ under different conditions when the spontaneous emission $\Gamma=0$.
          }
 \label{FNkr}
\end{figure}

In the above discussion, the dissipation has not been taken into account.
However, the system will interact with the environment inevitably which effects the
availability of this method. Thus, we investigate the
influence of spontaneous emission and photon leakage on this method. Once considered, the evolution of the system can be modeled by a master equation in Lindblad form
\begin{eqnarray}\label{eq23}
  \dot{\rho}=i[\rho,H_{tot}]+\sum_{k}{[L_{k}\rho L_{k}^{\dag}-\frac{1}{2}(L_{k}^{\dag}L_{k}\rho+\rho L_{k}^{\dag}L_{k})]},
\end{eqnarray}
where the $L_k$'s are the so-called Lindblad operators \cite{MJKFRASDPrl11}. The
five Lindblad operators governing dissipation in the two-atom model are
\begin{eqnarray}\label{eq24}
  L_{1}^{\kappa}=\sqrt{\kappa}a,\
  L_{2}^{\Gamma}=\sqrt{\Gamma_{1}}|f\rangle_{1}\langle e|,\
  L_{3}^{\Gamma}=\sqrt{\Gamma_{2}}|f\rangle_{2}\langle e|,\
  L_{4}^{\Gamma}=\sqrt{\Gamma_{3}}|g\rangle_{1}\langle e|,\
  L_{5}^{\Gamma}=\sqrt{\Gamma_{4}}|g\rangle_{2}\langle e|,
\end{eqnarray}
where $\kappa$ is the decay of the cavity and $\Gamma_{i}$
($i=1,2,3,4$) are the spontaneous emissions of atoms. Without loss
of generality, we set $\Gamma_{i}=\Gamma/2$. The fidelity $F$ for
the target state $|\psi_{5}\rangle$ is given through the relation
$F=|\langle\psi_{5}|\rho(t_{f})|\psi_{5}\rangle|$, where
$\rho(t_{f})$ is the density operator of the system at the time
$t_{f}$. In Fig. \ref{FNkr} (a) we plot
the fidelity $F$ of the target state $|\psi_{5}\rangle$ versus the
decay of spontaneous emission $\Gamma/\lambda$ with different values
of $\epsilon$ and $\lambda t_{f}$ when the decay of cavity
$\kappa/\lambda=0$. The result shows that the larger the value of
$\epsilon$ is, the more sensitive to the decay of spontaneous
emission the system is. The reason for this result is that the populations of effective intermediate states $|\psi_{2}\rangle$ and
$|\psi_{4}\rangle$ decrease as $\epsilon$ gets smaller. Figure \ref{FNkr} (b) shows the fidelity $F$ of the target state
$|\psi_{5}\rangle$ versus the decay of cavity $\kappa/\lambda$  with
different values of $\epsilon$ and $\lambda t_{f}$ when the decay of
spontaneous emission $\Gamma/\lambda=0$. The
sensitivity of the system to the decay of cavity seemingly decreases
with the decreasing of $\epsilon$. Because the population of
the effective intermediate state $|\psi_{3}\rangle$ is mainly
dominated by the ratio $r$, and the Zeno condition
($\lambda\gg\Omega_k$) could be satisfied very well when the ratio
$r$ is small enough, that is, the intermediate state
$|\psi_{3}\rangle$ can be effectively neglected with an adequately
small ratio $r$. Thus $r=0.3806$ when
$\epsilon=0.2636$ and $\lambda t_{f}=10$, $r=0.4195$ when
$\epsilon=0.1196$ and $\lambda t_{f}=20$, and $r=0.3273$ when
$\epsilon=0.0810$ and $\lambda t_{f}=40$. The population of
$|\psi_{3}\rangle$ is the smallest when $\epsilon=0.0810$ for the
three sets of parameters in Fig. \ref{FNkr}. As it is known, the interaction time $\lambda t_{f}$ also governs the
decoherence of the system. Considering both the
population of the state $|\psi_{3}\rangle$ and the interaction time
, the most insensitive to the decay of
cavity is at $\epsilon=0.2636$ and $\lambda t_{f}=10$. Contrast
Fig. \ref{FNkr} (a) with  Fig. \ref{FNkr} (b); an
increase in the decay rate $\kappa$ reduces the stationary state
fidelity more rapidly than an increase in the decay rate $\Gamma$.

\begin{figure}
 \renewcommand\figurename{\small FIG.}
 \centering \vspace*{8pt} \setlength{\baselineskip}{10pt}
 \subfigure[]{
 \includegraphics[scale = 0.33]{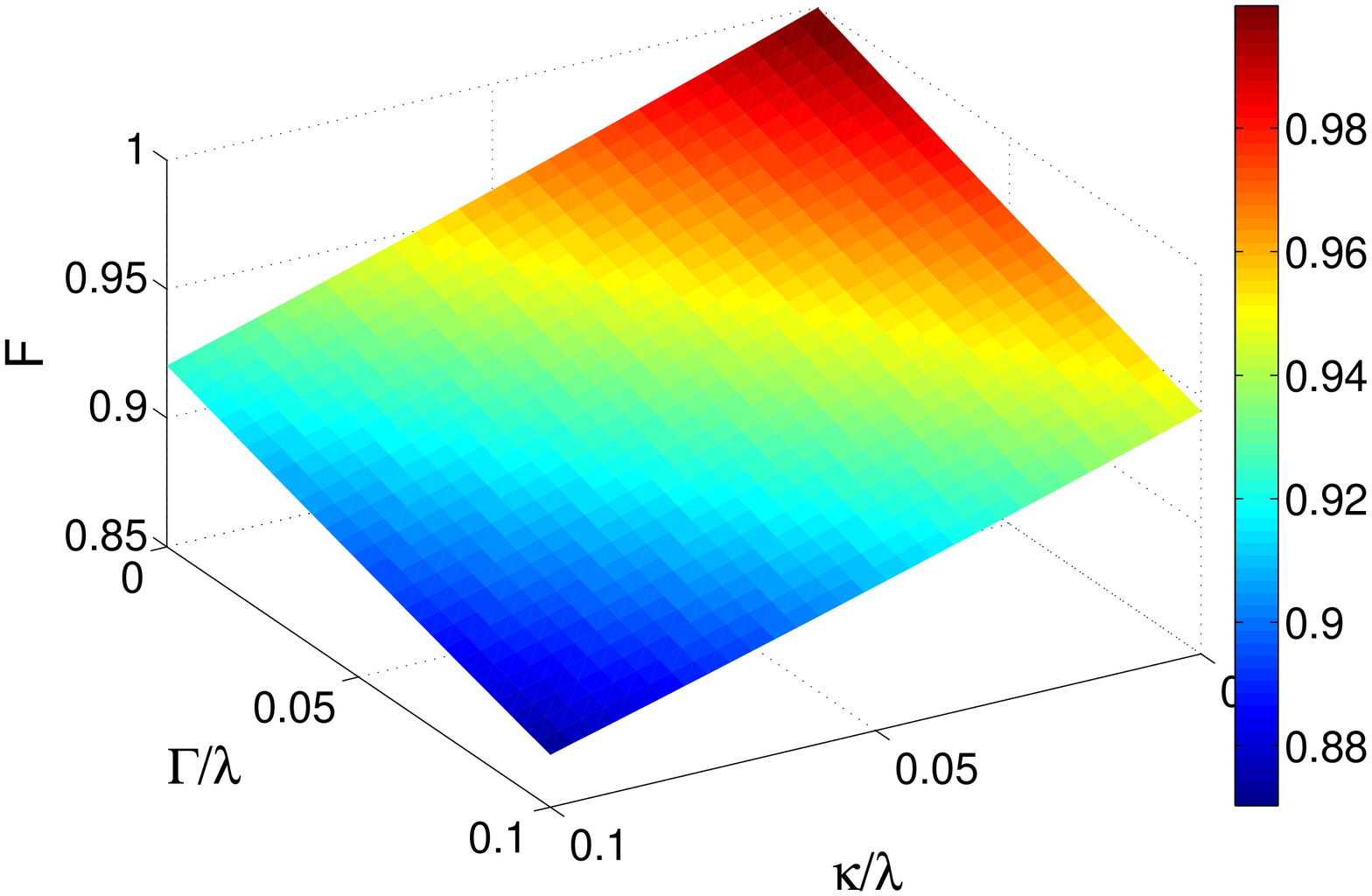}}
 \subfigure[]{
 \includegraphics[scale = 0.33]{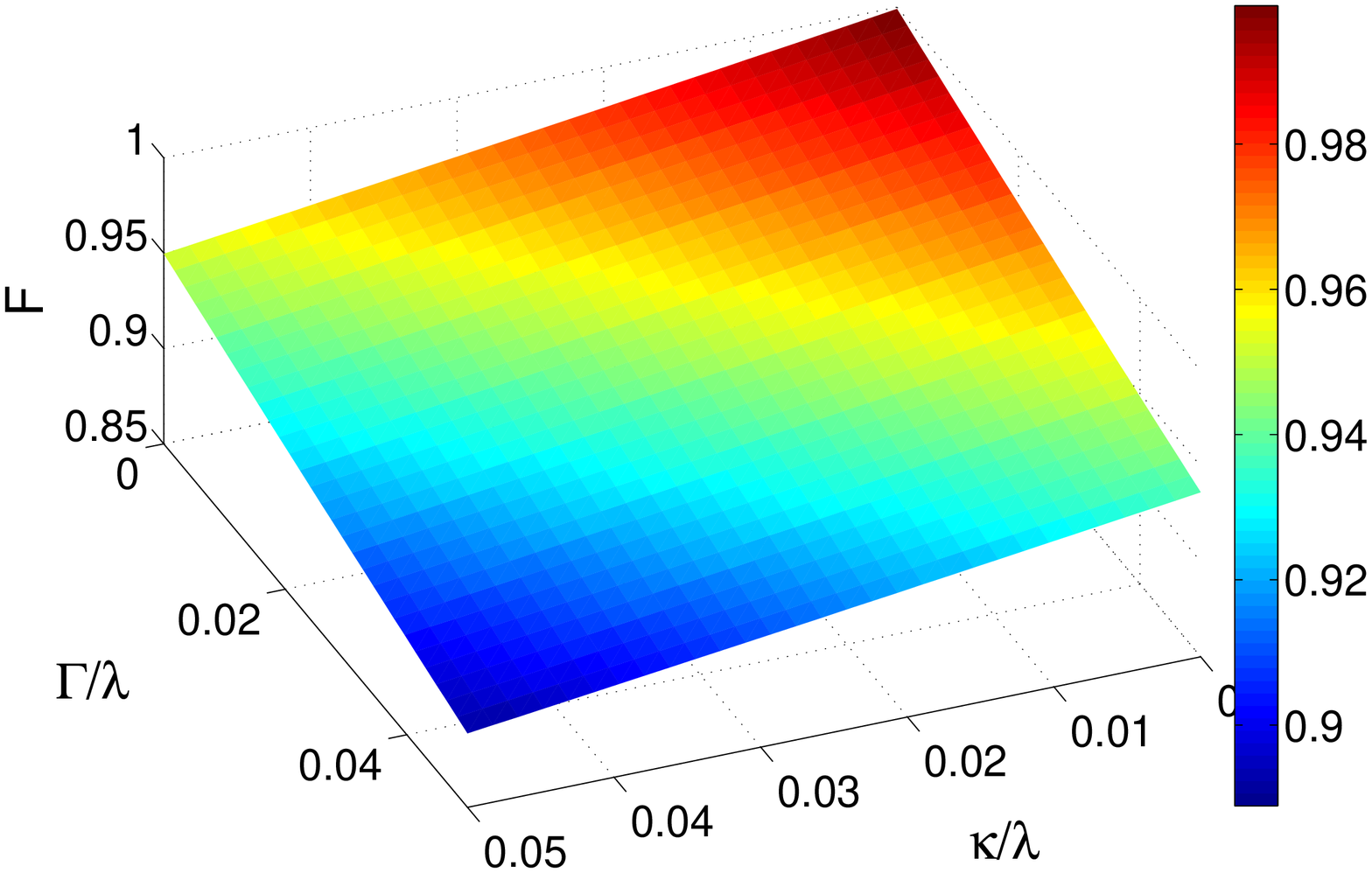}}
 \caption{
    (a) The fidelity $F$ of the target state $|\psi_5\rangle$ versus the ratios $\Gamma/\lambda$ and $\kappa/\lambda$ in the two-atom system.
    (b) The fidelity $F$ of the target state $|\varphi_7\rangle$ versus the ratios $\Gamma/\lambda$ and $\kappa/\lambda$ in the three-atom system.
          }
 \label{Fkr}
\end{figure}

The relationship of the fidelity $F$ of the target state
$|\psi_{5}\rangle$ versus the ratios $\kappa/\lambda$
and $\Gamma/\lambda$ by solving the master equation numerically
is shown in Fig. \ref{Fkr} (a) when $\epsilon=0.2636$ and $\lambda t_{f}=10$.
The fidelity $F$ decreases slowly with the
increasing of cavity decay and atomic spontaneous emission
and it is robust against to cavity decay and atomic spontaneous emission since it is still
about $87.03\%$ when $\kappa/\lambda=\Gamma/\lambda=0.1$. Therefore, our scheme is robust against the
two error sources and could acquire a better result in realistic conditions.

\section{Fast populations transfer in the multiparticle systems}
Actually, this method can be effectively applied to a multiparticle
system for achieving the FPTs, generating entangled states,
implementing phase gates, etc.. Assume that all of the
atoms are trapped in one cavity, in the interaction picture,
the Hamiltonian of a cavity-atom combined system can be described as
\begin{eqnarray}\label{eq25}
  H_i=H_{ac}+H_{al}+H_{aa},
\end{eqnarray}
where $H_{ac}$ is the Hamiltonian for the interaction between the atoms and the cavity,
$H_{al}$ is the Hamiltonian for the interaction between the atoms and the time-dependent laser pulses,
and $H_{aa}$ is the Hamiltonian for the direct interaction between the atoms. In a typical setup with neutral atoms at least
several microns apart direct interactions are negligible, $H_{aa}=0$. Just as QZD,
with the eigenvectors of $H_{ac}$, we rewrite the Hamiltonian $H_{al}$ and $H_{ac}$
as $H'_{al}$ and $H'_{ac}$, respectively. By solving the characteristic equation
of $H_{ac}$, a set of eigenvalues $\xi_{n}=\sum C_{n,m}\lambda_{m}$ is gained. Here $\lambda_{m}$ is the $m$th coupling constant
between the atoms and cavity. Setting $\lambda_{m}=\lambda$ for simplicity, we get a set of eigenvalues
$\xi_{n}=C'_{n}\lambda$. The Hamiltonian are given by
\begin{eqnarray}\label{eq26}
  H_i&=&H'_{ac}+H'_{al},                                                                \cr\cr
  H'_{ac}&=&\sum_{n}{\xi_{n}|\Phi_{n}\rangle\langle\Phi_{n}|},                          \cr\cr
  H'_{al}&=&\sum_{n,m,l}{b_{n,m,l}\Omega_{m}(t)|\Phi_{n}\rangle\langle\varphi_{l}|}+H.c.,
\end{eqnarray}
where $|\Phi_{n}\rangle$ is the $n$th eigenvector for the Hamiltonian $H_{ac}$, $\Omega_{m}(t)$ is the
$m$th Rabi frequency for the whole system, $|\varphi_{l}\rangle$ is the $l$th basis vector for the whole system,
and $b_{n,m,l}$ is the corresponding $\{n,m,l\}$th coefficient.
Almost the same as the transition between Eq. (\ref{eq4}) and Eq. (\ref{eq5}), we perform a unitary transformation
$U=e^{-iH'_{ac}t}$ on $H'_{al}$ under the condition $H'_{al}\ll H'_{ac}$. We find that the Hamiltonian becomes $H^{eff}_{al}$,
\begin{eqnarray}\label{eq27}
  H_{al}^{eff}=\sum_{n,m,l}{b_{n,m,l}\Omega_{m}(t)e^{i\xi_{n}t}|\Phi_{n}\rangle\langle\varphi_{l}|}+H.c..
\end{eqnarray}
Suppose that there are
$M$ different eigenvalues for the Hamiltonian $H_{ac}$, and the corresponding eigenvalues are
$0,\ \pm\lambda,\ \pm\sqrt{2}\lambda,\ \pm\sqrt{3}\lambda,\ \cdots$. By utilizing the analysis in section \textrm{II},
we consider the terms with $\xi_{n}=0$ as the main subsystem $S_{1}$ for the whole system,
and the terms with eigenvalues $\pm\lambda$ as the secondary subsystems $S^{+}_{2}$ and $S^{-}_{2}$, and so on.
Firstly, we design, by invariant-based inverse engineering, resonant laser pulses to perform a FPT in
the main subsystem $S_{1}$. Secondly, by setting some simple conditions, a part of the subsystems can be neglected since
the interaction between the states in each of these subsystems is far weaker than that in the main subsystem.
Introducing some special vectors (a part of these vectors can be
neglected all the time during the evolution of the whole system
and the rest of the vectors only have direct interaction with the vectors which are neglected),
we rewrite the total Hamiltonian and find out the dark state.

Next, the most important work is how to design and perform the FPT in the subsystem $S_1$.
From refs \cite{XCILARDGJGMPrl10,XCJGMPra12,MLLTSYXJS13}, we know that
it is very hard to directly design and perform the FPT in a
system which is more complicated than the three-level single-atom system.
It is best to perform an equivalent transformation to
make the subsystem $S_1$ become a system which can be considered as a
two-level or three-level single-atom system. And the part
of these operations for achieving the ``excited state''
of the ``two-level or three-level single-atom system''
can be finished based on the superposition principle and
Gram-Schmidt orthonormalization since all of the
states $|\Phi_{n}\rangle$ in this subsystem have the same eigenvalue $\xi=0$.
We cipher out the conditions for neglecting the special vectors.
The whole system is alike to
the two-atom system mentioned in Sec. \textrm{II},
and then the FPT in a multiparticle system can be effectively achieved.

\begin{figure}
 \renewcommand\figurename{\small FIG.}
 \centering \vspace*{8pt} \setlength{\baselineskip}{10pt}
 \subfigure[]{
 \includegraphics[scale = 0.33]{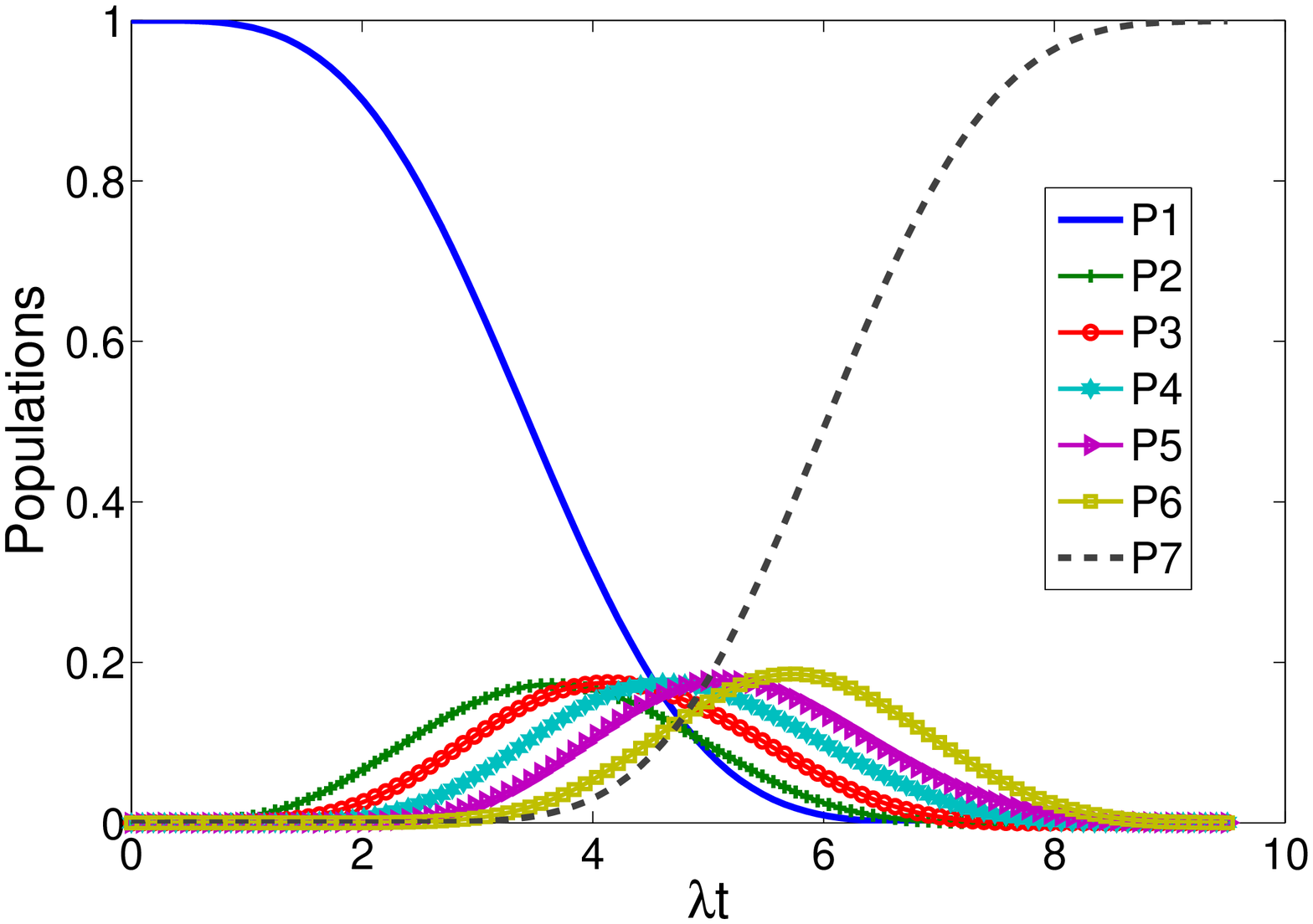}}
 \subfigure[]{
 \includegraphics[scale = 0.33]{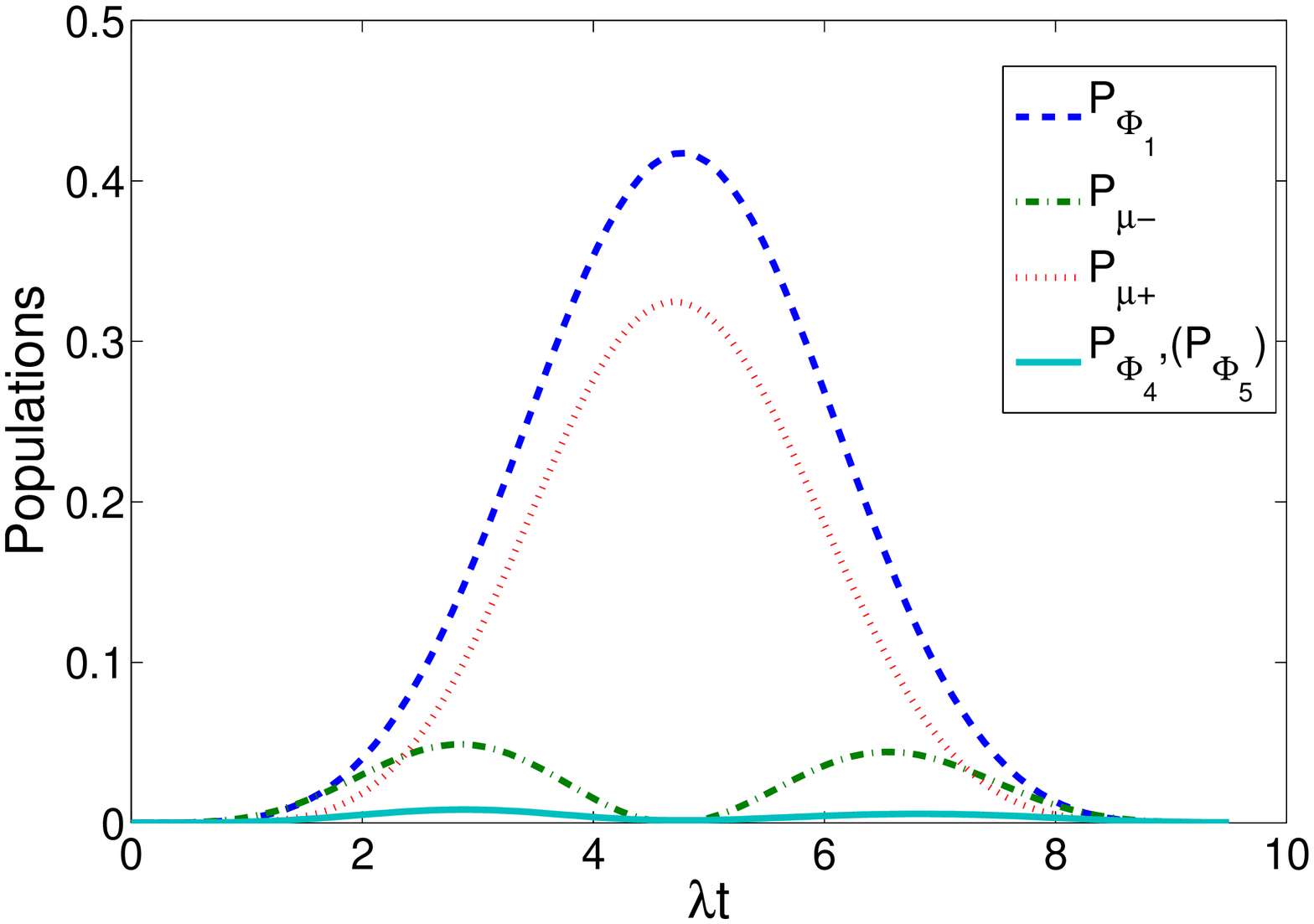}}
 \caption{
    (a) Time evolution of the populations for the states $|\varphi_{1}\rangle-|\varphi_{7}\rangle$ when $\epsilon=0.2596$ and $\lambda t_f=9.5$.
    (b) Time evolution of the populations for the states $|\Phi_{1}\rangle$, $|\mu_{+}\rangle$, $|\mu_{-}\rangle$, and $|\Phi_4\rangle$($|\Phi_5\rangle$)
        when $\epsilon=0.2596$ and $\lambda t_f=9.5$.
          }
 \label{PTatom}
\end{figure}

We now consider three atoms are trapped in a bimodal-mode cavity. Each
atom has one excited state $|e\rangle$ and three ground states $|f\rangle$, $|g_{+}\rangle$,
and $|g_{-}\rangle$. The transition $|f\rangle\leftrightarrow|e\rangle$ is resonantly driven
through a time-dependent laser pulse with Rabi frequency $\Omega(t)$, and the transition
$|g_{+}\rangle(|g_{-}\rangle)\leftrightarrow|e\rangle$ is resonantly coupled to the
left-circularly (right-circularly) polarized cavity mode with coupling constant $\lambda_{+}$($\lambda_{-}$).
The transition $|g_{+}\rangle_{1}(|g_{-}\rangle_{3})\leftrightarrow|e\rangle$ and $|f\rangle\leftrightarrow|e\rangle$ is supposed to be closed for atom $a_{1}$($a_{3}$) and atom $a_{2}$, respectively. As a consequence, the total Hamiltonian in the
interaction picture is given by
\begin{eqnarray}\label{eq28}
  H_{I}&=&H_{al}+H_{ac},                                              \cr\cr
  H_{al}&=&\sum_{k=1,3}{\Omega_{k}(t)|e\rangle_{k}\langle f|+H.c.},   \cr\cr
  H_{ac}&=&\lambda_{1,+}|e\rangle_{1}\langle g_{+}|a_{+}
           +\lambda_{2,+}|e\rangle_{2}\langle g_{+}|a_{+}
           +\lambda_{2,-}|e\rangle_{2}\langle g_{-}|a_{-}
           +\lambda_{3,-}|e\rangle_{3}\langle g_{-}|a_{-}
           +H.c.,
\end{eqnarray}
where the subscripts $1$, $2$, and $3$ represent the atoms $a_{1}$, $a_{2}$,
and $a_{3}$, respectively. $a_{\pm}$ are the annihilation operators for the cavity modes.
We assume the initial state is $|f\rangle_{1}|g_{+}\rangle_{2}|g_{-}\rangle_{3}|0\rangle_{c}$
and $\lambda_{k,\pm}=\lambda$ ($k=1,2,3$). The basis vectors for the whole system are
\begin{eqnarray}\label{eq31}
 |\varphi_1\rangle&=&|f\rangle_1|g_+\rangle_2|g_-\rangle_3|0\rangle_c,   \cr
 |\varphi_2\rangle&=&|e\rangle_1|g_+\rangle_2|g_-\rangle_3|0\rangle_c,   \cr
 |\varphi_3\rangle&=&|g_+\rangle_1|g_+\rangle_2|g_-\rangle_3|1\rangle_c, \cr
 |\varphi_4\rangle&=&|g_+\rangle_1|e\rangle_2|g_-\rangle_3|0\rangle_c,   \cr
 |\varphi_5\rangle&=&|g_+\rangle_1|g_-\rangle_2|g_-\rangle_3|1\rangle_c, \cr
 |\varphi_6\rangle&=&|g_+\rangle_1|g_-\rangle_2|e\rangle_3|0\rangle_c,   \cr
 |\varphi_7\rangle&=&|g_+\rangle_1|g_-\rangle_2|f\rangle_3|0\rangle_c,
\end{eqnarray}
and the eigenvectors for the Hamiltonian $H_{ac}$ are
\begin{eqnarray}\label{eq32}
 |\Phi_1\rangle&=&\frac{1}{\sqrt{3}}(|\varphi_2\rangle-|\varphi_4\rangle+|\varphi_6\rangle),                                                      \cr
 |\Phi_2\rangle&=&\frac{1}{2}(-|\varphi_2\rangle-|\varphi_3\rangle+|\varphi_5\rangle+|\varphi_6\rangle),                                          \cr
 |\Phi_3\rangle&=&\frac{1}{2}(-|\varphi_2\rangle+|\varphi_3\rangle-|\varphi_5\rangle+|\varphi_6\rangle),                                          \cr
 |\Phi_4\rangle&=&\frac{1}{2\sqrt{2}}(|\varphi_2\rangle+\sqrt{3}|\varphi_3\rangle+2|\varphi_4\rangle+\sqrt{3}|\varphi_5\rangle+|\varphi_6\rangle),\cr
 |\Phi_5\rangle&=&\frac{1}{2\sqrt{2}}(|\varphi_2\rangle-\sqrt{3}|\varphi_3\rangle+2|\varphi_4\rangle-\sqrt{3}|\varphi_5\rangle+|\varphi_6\rangle),
\end{eqnarray}
with  eigenvalues $\xi_1=0$, $\xi_2=\lambda$, $\xi_3=-\lambda$, $\xi_4=\sqrt{3}\lambda$, and $\xi_5=-\sqrt{3}\lambda$.
It is verified that the $H_{al}^{eff}$ is
\begin{eqnarray}\label{eq29}
  H_{al}^{eff}&=&\frac{1}{\sqrt{3}}|\Phi_{1}\rangle(\Omega_{1}(t)\langle\varphi_{1}|+\Omega_{3}(t)\langle\varphi_{7}|)                          \cr
               &&+\frac{1}{2}|\Phi_{2}\rangle(-\Omega_{1}(t)\langle\varphi_{1}|+\Omega_{3}(t)\langle\varphi_{7}|)e^{i\lambda t}                 \cr
               &&+\frac{1}{2}|\Phi_{3}\rangle(-\Omega_{1}(t)\langle\varphi_{1}|+\Omega_{3}(t)\langle\varphi_{7}|)e^{-i\lambda t}                \cr
               &&+\frac{1}{2\sqrt{3}}|\Phi_{4}\rangle(\Omega_{1}(t)\langle\varphi_{1}|+\Omega_{3}(t)\langle\varphi_{7}|)e^{i\sqrt{3}\lambda t}  \cr
               &&+\frac{1}{2\sqrt{3}}|\Phi_{5}\rangle(\Omega_{1}(t)\langle\varphi_{1}|+\Omega_{3}(t)\langle\varphi_{7}|)e^{-i\sqrt{3}\lambda t}
               +H.c..
\end{eqnarray}
Caused by five different eigenvalues of the Hamiltonian $H_{ac}$, we divide the system into five subsystems,
\begin{eqnarray}\label{eq30}
  S_{1}    &=&\{|\varphi_{1}\rangle,\ |\Phi_{1}\rangle,\ |\varphi_{7}\rangle\},     \
  S_{2}^{+} = \{|\varphi_{1}\rangle,\ |\Phi_{2}\rangle,\ |\varphi_{7}\rangle\},     \
  S_{2}^{-} = \{|\varphi_{1}\rangle,\ |\Phi_{3}\rangle,\ |\varphi_{7}\rangle\},     \cr
  S_{3}^{+}&=&\{|\varphi_{1}\rangle,\ |\Phi_{4}\rangle,\ |\varphi_{7}\rangle\},     \
  S_{3}^{-} = \{|\varphi_{1}\rangle,\ |\Phi_{5}\rangle,\ |\varphi_{7}\rangle\}.     \
\end{eqnarray}
The main subsystem $S_{1}$ can be considered as a
three-level single-atom system. If we set
``$\sqrt{2}\lambda$ is slightly larger than $\Omega_{k}$''
(actually, the setting varies depending on the method), the terms
containing the oscillating frequency $\pm\sqrt{3}\lambda$ will be
effectively neglected, that is, the subsystems $S_{3}^{\pm}$ can be
effectively neglected. The two vectors introduced for rewriting
the total Hamiltonian in eq. (\ref{eq28}) are
$|\mu_{+}\rangle=\frac{1}{\sqrt{2}}(|\Phi_{2}\rangle-|\Phi_{3}\rangle)$
and
$|\mu_{-}\rangle=\frac{1}{\sqrt{2}}(|\Phi_{2}\rangle+|\Phi_{3}\rangle)$.
The whole system evolves in the subspace spanned by the basis
vectors $\{|\varphi_{1}\rangle,\ |\varphi_{7}\rangle,\
|\Phi_{1}\rangle,\ |\mu_{+}\rangle,\ |\mu_{-}\rangle \}$. In terms of the basis
vectors, the total Hermitian in the interaction picture is simplified as
\begin{eqnarray}\label{eq34 H3re}
  H^{3}_{re}&=&\frac{1}{\sqrt{3}}|\Phi_{1}\rangle(\Omega_{1}(t)\langle\varphi_{1}|
                                               +\Omega_{3}(t)\langle\varphi_{7}|)   \cr
             &&+\frac{1}{\sqrt{2}}|\mu_{-}\rangle(-\Omega_{1}(t)\langle\varphi_{1}|+\Omega_{3}(t)\langle\varphi_{7}|)
               +\lambda|\mu_{-}\rangle\langle\mu_{+}|+H.c..
\end{eqnarray}
An ideal FPT is performed effectively in the whole system, and it is the same as what we have done in section \textrm{II}. First, we design the two special Rabi frequencies by using the dynamics of invariant-based inverse engineering.
The Hermitian $H^{3}_{S_{1}}$ for the main subsystem $S_{1}$ of the three-atom model reads
\begin{eqnarray}\label{eq34 Hs1}
  H^{3}_{S_1}(t)=\frac{1}{\sqrt{3}}\left(
                                    \begin{array}{ccc}
                                        0 & \Omega_{1}(t) & 0             \\
                                        \Omega_{1}(t) & 0 & \Omega_{3}(t) \\
                                        0 & \Omega_{3}(t) & 0             \\
                                    \end{array}
                               \right).
\end{eqnarray}
And the corresponding invariant Hermitian operator $I^{3}_{S_{1}}(t)$ satisfying
$i\partial I^{3}_{S_{1}}(t)/\partial t=[H^{3}_{S_{1}}(t),I^{3}_{S_{1}}(t)]$ for speeding up the transfer is
\begin{eqnarray}\label{eq35 Is1}
   I^{3}_{S_{1}}(t)=\frac{1}{\sqrt{3}}\chi'
              \left(
                \begin{array}{ccc}
                 0 & \cos{\gamma'}\sin{\beta'} & -i\sin{\gamma'}            \\
                 \cos{\gamma'}\sin{\beta'} & 0 & \cos{\gamma'}\cos{\beta'}  \\
                 i\sin{\gamma'} & \cos{\gamma'}\cos{\beta'} & 0             \\
              \end{array}
             \right),
\end{eqnarray}
where $\chi'$ is an arbitrary constant with units of frequency to keep $I^{3}_{S_{1}}(t)$ involving the energy dimension.
Then the two special Rabi frequencies designed for performing the FPT in the main subsystem $S_{1}$ are inferred,
\begin{eqnarray}\label{eq36 omega1omega2}
  \Omega_{1}(t)=\sqrt{3}(\dot{\beta'}\cot{\gamma'}\sin{\beta'}+\dot{\gamma'}\cos{\beta'}), \cr
  \Omega_{3}(t)=\sqrt{3}(\dot{\beta'}\cot{\gamma'}\cos{\beta'}-\dot{\gamma'}\sin{\beta'}).
\end{eqnarray}
We also choose $\gamma'=\epsilon$ and $\beta'=\pi t/t_{f}$, where $\epsilon$ is also a small
value which should be carefully chosen later for a high fidelity of the transfer. Substituting $\gamma'$ and $\beta'$ into eq.(\ref{eq36 omega1omega2}), the two special Rabi frequencies
turnout to be
\begin{eqnarray}\label{eq37 omega1omega3}
  \Omega_{1}(t)=(\sqrt{3}\pi/2t_{f})\cot{\epsilon}\sin(\pi t/2t_{f}),\cr
  \Omega_{3}(t)=(\sqrt{3}\pi/2t_{f})\cot{\epsilon}\cos(\pi t/2t_{f}).
\end{eqnarray}
Second, we make the secondary subsystems $S^{\pm}_{2}$ become the auxiliary for the FPT in the whole system.
By solving the intrinsic equation of $H^{3}_{re}$, the dark state for the whole system is obtained,
\begin{eqnarray}\label{eq34}
  |Dark_{3}\rangle&=&\frac{1}{\sqrt{N_{3}}}[\Omega_{3}(t)|\varphi_{1}\rangle
                                          -\frac{\Omega_{1}(t)\Omega_{3}(t)}{\lambda}(|\varphi_{3}\rangle-|\varphi_{5}\rangle)
                                          -\Omega_{1}(t)|\varphi_{7}\rangle]              \cr\cr
                  &=&\frac{1}{\sqrt{N_{3}}}[\Omega_{3}(t)|\varphi_{1}\rangle
                                          +\frac{\Omega_{1}(t)\Omega_{3}(t)}{\lambda}(|\Phi_{2}\rangle-|\Phi_{3}\rangle)
                                          -\Omega_{1}(t)|\varphi_{7}\rangle],             \cr\cr
                  &=&\frac{1}{\sqrt{N_{3}}}[\Omega_{3}(t)|\varphi_{1}\rangle
                                          +\frac{\sqrt{2}\Omega_{1}(t)\Omega_{3}(t)}{\lambda}|\mu_{+}\rangle
                                          -\Omega_{1}(t)|\varphi_{7}\rangle],
\end{eqnarray}
with $N_{3}=\sqrt{\Omega_{1}(t)^{2}+\Omega_{3}(t)^{2}+2(\Omega_{1}(t)\Omega_{3}(t)/\lambda)^2}$.
The intermediate state $|\mu_{-}\rangle$ is considered as a state which can be
neglected all the time and the state $|\mu_{+}\rangle$ is considered as an
independent state of the system under certain conditions. By setting the
condition for very slightly increasing the population of $|\mu_{-}\rangle$,
the FPT of the whole system can be achieved.
And a very short interaction time i.e., $\lambda t_{f}=9.5$, is needed
for achieving a perfect target state $|\varphi_{7}\rangle$ with a fidelity
$99.9\%$ from the initial state $|\varphi_1\rangle$ when $\epsilon=0.2596$
by the numerical calculation. Figure \ref{PTatom} (a) shows the time evolution
of the populations for states $|\varphi_{1}\rangle-|\varphi_{7}\rangle$. Figure \ref{PTatom} (b) is plotted to demonstrate that the subsystems
$S_{3}^{\pm}$ and the state $|\mu_{-}\rangle$ can be effectively neglected.
From Fig. \ref{PTatom} (b), it is displayed that the populations of the
states $|\Phi_{4}\rangle$ and $|\Phi_{5}\rangle$ remain negligible all the time
since the maximum values of the populations are only $0.82\%$ for the states
$|\Phi_{4}\rangle$ and $|\Phi_{5}\rangle$. The state $|\mu_{-}\rangle$ is
very slightly populated for speeding up the population transfer, and it still can be considered as
negligible since the maximum value of its population is only $4.8\%$.
Figs. \ref{PTatom} (a) and (b) are plotted when $\epsilon=0.2596$ and $\lambda t_{f}=9.5$. The fidelity
of the target state $|\varphi_{7}\rangle$ in the presence of decoherence is
given through solving the master equation according to Eq. (\ref{eq23}).
There are eight Lindblad operators for the three-atom model,
\begin{eqnarray}\label{eq33}
  L_{1}^{\kappa}&=&\sqrt{\kappa_{+}}a_{+},                      \
  L_{2}^{\kappa} = \sqrt{\kappa_{-}}a_{-},                      \
  L_{3}^{\Gamma} = \sqrt{\Gamma_{1}}|f\rangle_{1}\langle e|,    \
  L_{4}^{\Gamma} = \sqrt{\Gamma_{2}}|f\rangle_{3}\langle e|,    \cr
  L_{5}^{\Gamma}&=&\sqrt{\Gamma_{3}}|g_{+}\rangle_{1}\langle e|,\
  L_{6}^{\Gamma} = \sqrt{\Gamma_{4}}|g_{-}\rangle_{3}\langle e|,\
  L_{7}^{\Gamma} = \sqrt{\Gamma_{5}}|g_{+}\rangle_{2}\langle e|,\
  L_{8}^{\Gamma} = \sqrt{\Gamma_{6}}|g_{-}\rangle_{2}\langle e|.\
\end{eqnarray}
We also set $\kappa_{i}=\kappa$ ($i=+,-$) and $\Gamma_{j}=\Gamma/2$ ($j=1,2,\cdots,6$) for simplicity.
From the relationship of fidelity $F$ of the target
state $|\varphi_{7}\rangle$ versus the ratios $\kappa/\lambda$
and $\Gamma/\lambda$ given in FIG. \ref{Fkr} (b), $F$ decreases slowly with the increasing of cavity decay and
atomic spontaneous emission and it is insensitive to both of these
two error sources because it is still about $88.89\%$ when $\kappa=\Gamma=0.05\lambda$.

\section{CONCLUSION}
The invariant-based inverse method presented here may be compared to
the optimal control approaches in Refs.
\cite{XCJGMPra12,VSMDJTPra97,IRSVSMDJT,UBGCJPGSGHRJJmp03}, it
provides a complementary perspective of these approaches, whereas
optimal control is useful to choose among the possible solutions
found by the invariant-based inverse engineering \cite{XCJGMPra12}.
The QZD is a very effective method and it has been widely used in
quantum information processing
\cite{ABDBPLKNjp00,HAPra03,RXCLTSPla,XBWJQYFNPra08}. It is well
known that the QZD has the advantage of simplifying a complicated
system by space division, and the shortcuts to
adiabatic passage mentioned by Chen \emph{et al.} has
the advantage of shortening the operation time by using special
resonant pulses. In this paper, we combine the advantage of
``simplifying a complicated system'' with the advantage of
``shortening the operation time'', and present a method for
performing the FPTs in multiparticle systems. Two different models
have been discussed, and a perfect target state can be achieved in a
very short interaction time in each of the two models. But some
relatively large laser intensities are needed since shortening the
time implies an energy cost \cite{XCJGMPra12}. In a more general
case, if there are no eigenvalues $\xi_{n}=0$ for the Hamiltonian
$H_{ac}$, the Hamiltonian for the main subsystem $H_{S_{1}}$ does
not possess SU(2) symmetry, so that the invariant $I_{S_{1}}$ should
be constructed in terms of the eight Gell-Mann matrices for the
SU(3) group \cite{FTHPra85}.

In experiment, the atom cesium can be used for this method. And a set of cavity QED parameters
$(\lambda,\ \kappa,\ \Gamma)/2\pi=(750,\ 3.5,\ 2.62)$ MHz
is predicted to be available in an optical cavity \cite{SMSTJKKJVKWGEWHJKPra05},
therefore, the fidelity for the target state is still higher than
$99.2\%$ for the two-atom system. With these parameters, it allows us to
construct an atomic system for the FPT in the presence of decoherence.

In summary, we have proposed a promising method to construct
shortcuts to perform the FPT for ground states in two or more atoms
systems by invariant-based inverse engineering and in the view of
quantum Zeno dynamics in the cavity QED system. Compared with the
previous works, the present work can perform perfect FPTs in
multiparticle systems without additional complex conditions. And
this method is insensitive to the variations of the parameters, at
the same time, the interaction time needs not to be controlled
accurately. We firmly believe that this work will make contributions
to quantum information processing including performing atomic
transport, implementing quantum gates, generating entangled states,
etc..

\section{Acknowledgments}

This work was supported by the National Natural Science Foundation
of China under Grants No. 11205037 and No. 11105030, the Major State
Basic Research Development Program of China under Grant No.
2012CB921601,

\newpage


\end{document}